\documentclass[12pt]{article}
\usepackage{latexsym,cite,subeqn}
\input amssym.def
\input amssym.tex

\makeatletter  
\renewcommand{\theequation}{\thesection.\arabic{equation}}
\@addtoreset{equation}{section}
\makeatother

\font\msym=msbm10

\def\Real{{\mathop{\hbox{\msym \char  '122}}}} 
 
\def\cQ{{\cal Q}}
\def\Z{{\cal Z }}

\def\x{{\rm x}}

\def\a{{\rm a}}

\def\bt{\bar{\theta}}
\def\bT{\bar{\Theta}}

\def\X{{\rm X}}

\def\N{{\cal N}}

\def\e{\epsilon}
\def\h{{\rm h}}
\def\v{{\rm v}}

\begin{document}
\begin{titlepage}
\title{\vskip -60pt
{\small
\begin{flushright} 
KIAS-99101\\
hep-th/9910199
\end{flushright}}
\vskip 20pt
Superconformal Symmetry in Three-dimensions \\ ~~~\\~~~ }
\author{Jeong-Hyuck Park\thanks{E-mail address:\,\,jhp@kias.re.kr}}
\date{}
\maketitle
\vspace{-1.0cm}
\begin{center}
\textit{School of Physics, Korea Institute for Advanced Study}\\
\textit{207-43 Cheongryangri-dong, Dongdaemun-gu}\\
\textit{Seoul 130-012, Korea}\\
~~~\\
~~~\\
~~~\\
\end{center}
\begin{abstract}
\noindent Three-dimensional   ${\cal N}$-extended superconformal  
symmetry is  studied  within the superspace formalism. 
A superconformal Killing  equation is derived and  its  solutions are  
classified in terms of   supertranslations, dilations, Lorentz
transformations, $R$-symmetry transformations and special superconformal 
transformations.   Superconformal  group is then  identified with  a  
supermatrix group, $\mbox{OSp}(\N|2,\Real)$,  as expected from the 
analysis on  simple Lie superalgebras.   
In general, due to  the invariance under  supertranslations and special 
superconformal transformations,  superconformally  invariant   $n$-point  
functions  reduce to one  unspecified $(n-2)$-point function 
which must transform  homogeneously  under the 
remaining rigid transformations, i.e. dilations,  Lorentz    
transformations and  $R$-symmetry transformations.   
After constructing building blocks for superconformal   
correlators,  we are able to identify all the superconformal invariants and 
obtain the general form of $n$-point functions.  
Superconformally covariant differential operators are also discussed. 
\end{abstract}
~\\
{\small
\begin{flushleft} 
\end{flushleft}}
\thispagestyle{empty}
\end{titlepage}
\newpage

\section{Introduction and  Summary\label{intro}}
Based on the classification of  simple Lie superalgebras~\cite{kac}, 
Nahm analyzed  all possible superconformal algebras~\cite{nahm}. 
According to Ref. \cite{nahm},  
not  all spacetime dimensions allow the corresponding 
supersymmetry algebra to be extended to a superconformal algebra contrary to  
the ordinary conformal symmetry.  The standard supersymmetry algebra 
admits an extension to a superconformal algebra only 
if $d\leq 6$. Namely the highest dimension admitting superconformal algebra 
is six, and in $d=3,4,5,6$ dimensions      
the bosonic part of the superconformal algebra has the form
\begin{equation}
{\cal L}_{C}\oplus{\cal L}_{R}\,,
\end{equation}
where ${\cal L}_{C}$ is the Lie algebra of the conformal group and
${\cal L}_{R}$ is a $R$-symmetry algebra acting on the superspace  
Grassmann variables.  \newline
Explicitly for Minkowskian spacetime
\begin{equation}
\begin{array}{l}
d=3;~~~~~\mbox{o}(2,3)\oplus\mbox{o}(\N)\,,\\
{}\\
d=4;~\,\left\{\begin{array}{l}
\mbox{o}(2,4)\oplus\mbox{u}(\N)\,,~~\N\neq 4\\
{}\\
\mbox{o}(2,4)\oplus\mbox{su}(4)
\end{array}\right.\,,\\
{}\\
d=5;~~~~~\mbox{o}(2,5)\oplus\mbox{su}(2)\,,\\
{}\\
d=6;~~~~~\mbox{o}(2,6)\oplus\mbox{sp}(\N)\,,
\end{array}
\label{bosonicpart}
\end{equation}
where  $\N$ or the number appearing in $R$-symmetry part is related to 
the number of supercharges.\newline

\indent On six-dimensional Minkowskian spacetime it is possible to define 
Weyl spinors of opposite chiralities   
and so the general six-dimensional supersymmetry may be 
 denoted by two numbers, $(\N,\tilde{\N})$, 
where $\N$ and $\tilde{\N}$ are the numbers of chiral and
anti-chiral supercharges. The $R$-symmetry group is then 
$\mbox{Sp}(\N)\times\mbox{Sp}(\tilde{\N})$. The analysis of 
Nahm  shows that to admit a superconformal  algebra
either $\N$ or $\tilde{\N}$ should be zero. Although both 
 $(1,1)$ and $(2,0)$ supersymmetry give rise  $\N=4$ 
four-dimensional supersymmetry after dimensional reduction, only $(2,0)$
supersymmetry theories can be superconformal~\cite{seiberg16}.  
On five-dimensional  Minkowskian spacetime  Nahm's analysis seems to imply  
a certain   restriction on the  number of supercharges as the corresponding  
$R$-symmetry algebra is to be $\mbox{su}(2)$.   \newline

\indent   The above analysis  is essentially based on the 
classification of  simple Lie superalgebras and identification  of
the  bosonic part with  the  usual spacetime conformal symmetry rather
than Poincar\'{e} symmetry, since 
the  former  forms a simple group, while the latter does not. 
This approach does not rely on  any definition of 
superconformal transformations on superspace. \newline


\indent   The present paper deals with superconformal symmetry 
in three-dimensions and lies in the same framework   as 
our sequent work on superconformal symmetry in other  
dimensions, $d=4,\,6$  \cite{paper3,paper1,paper2}.  Namely we analyze   
superconformal symmetry directly in terms of       
coordinate transformations on  superspace.   We first define the  
superconformal group  on superspace and derive the superconformal Killing 
equation.   Its general solutions  are  identified in terms of  
supertranslations, dilations, Lorentz transformations, 
$R$-symmetry transformations and special superconformal 
transformations.  Based on the explicit form of the solutions 
the superconformal group is 
independently identified to agree with Nahm's analysis   
and some representations are obtained. \newline

\indent Specifically, in Ref. \cite{paper3} we identified four-dimensional 
$\N\neq 4$  extended superconformal group with  a  
supermatrix group, $\mbox{SU}(2,2|{\cal N})$, having  dimensions 
$(15+{\cal N}^{2}|8{\cal N})$, while for ${\cal N}=4$ case   we pointed out 
that  an equivalence relation must be imposed on the 
supermatrix group and so the four-dimensional ${\cal N}=4$ superconformal 
group is isomorphic to a  quotient group of the  supermatrix group. In fact  
${\cal N}=4$ superconformal group is a semi-direct product  of $\mbox{U}(1)$ 
and a simple Lie supergroup containing $\mbox{SU}(4)$.    
The $\mbox{U}(1)$ factor can be removed by imposing tracelessness condition on 
the supermatrix group so that the dimension reduces from $(31|32)$ to 
$(30|32)$  and the $R$-symmetry group  shrinks from   
$\mbox{U}(4)$ to Nahm's result,  $\mbox{SU}(4)$\footnote{Similarly if five-dimensional superconformal group is not simple,  this will be a  way out from the  
puzzling restriction on the  number of supercharges  in five-dimensional   
superconformal theories, as the corresponding   $R$-symmetry group can be bigger than Nahm's result, $\mbox{SU}(2)$.  However this is at the level of speculation at present.  }.  
In Ref. \cite{paper2} by solving the superconformal 
Killing equation we show that  six-dimensional $(\N,0)$ superconformal 
group is identified with a supermatrix group, $\mbox{OSp}(2,6|\N)$, having 
dimensions  $(28+\N(2\N+1)|16{\cal N})$,  while 
for $(\N,\tilde{\N})$, $\N,\tilde{\N}> 0$  supersymmetry,  we verified that 
although dilations may be introduced,  there exist no special superconformal 
transformations as expected from  Nahm's result.   \newline

\indent The main advantage of  our formalism is that it enables us to write  
general expression  for two-point, three-point and $n$-point 
correlation functions  of quasi-primary superfields which 
transform simply under superconformal transformations.    
In Refs. \cite{paper3,paper1,paper2}  we explicitly constructed  building 
blocks for  superconformal correlators  in four-  and six-dimensions,  and 
proved that these  building blocks actually  generate  
the general   form of  correlation functions.      
In general, due to  the  invariance under    supertranslations and special 
superconformal transformations,  
$n$-point functions  reduce to one  unspecified $(n-2)$-point function 
which must transform  homogeneously  under the 
rigid transformations only - dilations, Lorentz 
transformations and  $R$-symmetry transformations~\cite{paper3}.     
This feature of superconformally invariant correlation functions is universal  
for any spacetime dimension if  there exists  a well defined  
superinversion in the corresponding dimension, since  
superinversion plays a crucial role  in its proof.  For non-supersymmetric case, contrary to superinversion,   
the inversion map is  defined of the same form irrespective of the 
spacetime dimension and  hence   $n$-point functions  reduce to one  unspecified $(n-2)$-point function    in any dimension   which transform  homogeneously  under  dilations and  Lorentz transformations. \newline

\indent The formalism is powerful for applications whenever  
there exist off-shell superfield formulations for superconformal 
theories, and  such formulations are known in four-dimensions for 
$\N=1,2,3$ \cite{sohnius,wessbagger,buchbinder,peterwest,CQG2155}  
and in three-dimensions for $\N=1,2,3,4$ \cite{thousand,PLB243105,PLB268203,PLB28172,IJMPA8193371,CQG693,9804167,9902038}. 
In fact within the formalism  Osborn elaborated   the analysis of   
$\N=1$ superconformal symmetry for four-dimensional  quantum field 
theories~\cite{9808041}, and recently  Kuzenko and Theisen determine   
the general  structure of two- and three- point functions of the supercurrent  
and the flavour current of $\N=2$ superconformal field  
theories~\cite{9907107}.    A common result contained in  
Refs. \cite{9808041,9907107} is that  
the three-point functions of the  conserved 
supercurrents in both $\N=1$ and $\N=2$ superconformal theories 
allow two linearly independent structures and hence there exist 
two numerical coefficients which can  
be calculated in  specific perturbation theories using 
supergraph techniques.    \newline

\indent  The contents of the present paper are as follows.  
In section~\ref{preliminary} we review supersymmetry in three-dimensions. 
In particular, we verify that 
supersymmetry algebra with $N$ Dirac supercharges is 
equivalent to $2N$-extended Majorana supersymmetry algebra, so that in the 
present paper we consider  $\N$-extended Majorana superconformal symmetry 
with an arbitrary natural number, $\N$. \newline

\indent  In section~\ref{scfs}, 
we  first define the three-dimensional ${\cal N}$-extended  
superconformal group in terms of 
coordinate transformations on superspace as  a generalization
of the definition of ordinary conformal transformations. We then
derive a superconformal Killing equation,  
which is a necessary and sufficient  condition for a supercoordinate 
transformation to be superconformal.  
The general solutions are  identified in terms of  
supertranslations, dilations, Lorentz 
transformations, $R$-symmetry transformations and special superconformal 
transformations, where $R$-symmetry is given by  $\mbox{O}({\cal N})$ 
as in eq.(\ref{bosonicpart}).  We also  present  a  definition of 
superinversion in three-dimensions through which  
supertranslations and special 
superconformal transformations are dual to each other.  
The three-dimensional ${\cal N}$-extended  superconformal 
group is then  identified with  a  
supermatrix group, $\mbox{OSp}(\N|2,\Real)$, having  dimensions 
$(10+\textstyle{\frac{1}{2}}\N(\N-1)|4{\cal N})$ as expected from 
the analysis on  simple Lie superalgebras~\cite{nahm,9910030}.  \newline

\indent In section~\ref{coset}, we obtain an explicit formula for
the finite non-linear 
superconformal transformations of the supercoordinates, $z$,
parameterizing superspace and discuss several 
representations of the superconformal group.   
We also construct matrix or vector valued 
functions depending on  two or   three points 
in superspace which transform 
covariantly under superconformal transformations.  
For two points, $z_{1}$ and $z_{2}$, we find a matrix, 
$I(z_{1},z_{2})$, which transforms covariantly  like a
product of two tensors at $z_{1}$ and $z_{2}$.   
For three points, $z_{1},z_{2},z_{3}$, we find `tangent' vectors, 
$Z_{i}$, which transform homogeneously at $z_{i},\,i=1,2,3$. 
These  variables serve as  building blocks of     
obtaining two-point, three-point and general $n$-point 
correlation functions later. \newline

\indent In section~\ref{correlation}, we discuss the    
superconformal invariance of correlation 
functions for  quasi-primary superfields and exhibit general forms of   
two-point, three-point and $n$-point functions. 
Explicit formulae for two-point  functions of   
superfields in various cases are given.    
We   also identify all the superconformal invariants. \newline

\indent In section \ref{operator},  superconformally  
covariant differential operators are  discussed.  The 
conditions for superfields, which are  formed by the action of  
spinor derivatives on quasi-primary superfields, to remain 
quasi-primary are obtained.  In general, the action of 
differential operator on quasi-primary fields generates an anomalous term 
under superconformal transformations. However, with a suitable choice of 
scale dimension, we show that the anomalous term may be cancelled. 
We regard this analysis as a necessary step to  write superconformally 
invariant actions on superspace,  as the kinetic terms in such theories may 
consist of superfields formed by the action of  
spinor derivatives on quasi-primary superfields.  \newline

\indent   In the appendix,  the explicit form of superconformal algebra and a 
method of solving the superconformal Killing equation are exhibited. \newline

\newpage

\section{Preliminary\label{preliminary}}
\subsection{Gamma Matrices}
With the three-dimensional  Minkowskian metric, $\eta^{\mu\nu} =
\mbox{diag}(+1,-1,-1)$, the  $2\times 2$ gamma matrices,
$\gamma^{\mu},~\mu=0,1,2$,   
satisfy 
\begin{equation}
\gamma^{\mu}\gamma^{\nu}=\eta^{\mu\nu}+i\epsilon^{\mu\nu\rho}\gamma_{\rho}\,.
\label{gg}
\end{equation}
The hermiticity condition is 
\begin{equation}
\gamma^{0}\gamma^{\mu}\gamma^{0}=\gamma^{\mu}{}^{\dagger}\,.
\label{dagger}
\end{equation}
Charge conjugation matrix, $\e$, satisfies\footnote{To emphasize  
the anti-symmetric property of the $2\times 2$  charge conjugation matrix in three-dimensions we adopt the symbol, $\e$, instead of the conventional one, $C$.}~\cite{kugotownsend}
\begin{equation}
\begin{array}{cc}
\multicolumn{2}{c}{\e\gamma^{\mu}\e^{-1}=-\gamma^{\mu}{}^{t}\,,}\\
{}&{}\\
\e^{t}=-\e\,,~~~~&~~~~\e^{\dagger}=\e^{-1}\,.
\end{array}
\label{charge}
\end{equation}

\indent $\gamma^{\mu}$ forms a basis for $2\times 2$  traceless matrices with the completeness relation
\begin{equation}
\gamma^{\mu}{}^{\alpha}{}_{\beta}\gamma_{\mu}{}^{\gamma}{}_{\delta}=
2\delta^{\alpha}{}_{\delta}\delta_{\beta}{}^{\gamma}-\delta^{\alpha}{}_{\beta}
\delta^{\gamma}{}_{\delta}\,.
\label{complete}
\end{equation}

\subsection{Three-dimensional Superspace}
The three-dimensional  
supersymmetry algebra has the standard  form 
with $P_{\mu}=(H,-\textbf{P})$ 
\begin{equation}
\begin{array}{c}
\{\cQ^{i\alpha}, \bar{\cQ}_{j\beta}\}
=2\delta^{i}{}_{j}\gamma^{\mu}{}^{\alpha}{}_{\beta}P_{\mu}\,,\\
{}\\
{}[{P}_{\mu},{P}_{\nu}]=[{P}_{\mu},\cQ^{i\alpha}]=
[P_{\mu},\bar{\cQ}_{i\alpha}]=\{\cQ^{i\alpha},\cQ^{j\beta}\}=
\{\bar{\cQ}_{i\alpha},\bar{\cQ}_{j\beta}\}=0\,,
\end{array}
\label{3Dsusypre}
\end{equation}
where $1\leq \alpha\leq 2$,\,   $1\leq i\leq N$ and  
$\cQ^{i},\bar{\cQ}_{j}$  satisfy
\begin{equation}
\bar{\cQ}_{i}=\cQ^{i\dagger}\gamma^{0}\,.
\label{Qbar}
\end{equation}

Now we define for $1\leq a\leq 2N,\,1\leq i,j\leq N$
\begin{equation}
Q^{a}=\left(\begin{array}{c}
\textstyle{\frac{1}{\sqrt{2}}}
(\cQ^{i}+\e^{-1}\bar{\cQ}_{i}^{t})\\
{}\\
i\textstyle{\frac{1}{\sqrt{2}}}
(\cQ^{j}-\e^{-1}\bar{\cQ}_{j}^{t})
\end{array}\right)\,,
\label{extend1}
\end{equation}
and
\begin{equation}
\bar{Q}_{a}\equiv Q^{a\dagger}\gamma^{0}=
\left(\textstyle{\frac{1}{\sqrt{2}}}(\bar{\cQ}_{i}-\cQ^{it}\e),~~
-i\textstyle{\frac{1}{\sqrt{2}}}(\bar{\cQ}_{j}+\cQ^{jt}\e)\right)\,.
\end{equation}
$Q^{a},\bar{Q}_{b}$ satisfy the Majorana condition
\begin{equation}
\begin{array}{c}
\bar{Q}_{a}=Q^{a\dagger}\gamma^{0}=-Q^{a}{}^{t}\e\,,\\
{}\\
Q^{a}=\e^{-1}\bar{Q}_{a}^{t}\,.
\end{array}
\label{Majorana}
\end{equation}
With this notation we note that the  three-dimensional $N$-extended 
supersymmetry algebra~(\ref{3Dsusypre}) is equivalent to the $2N$-extended Majorana  supersymmetry algebra
\begin{equation}
\begin{array}{c}
\{Q^{a\alpha}, \bar{Q}_{b\beta}\}
=2\delta^{a}{}_{b}\gamma^{\mu}{}^{\alpha}{}_{\beta}P_{\mu}\,,\\
{}\\
{}[{P}_{\mu},{P}_{\nu}]=[{P}_{\mu},Q^{a\alpha}]=0\,.
\end{array}
\label{3Dsusy}
\end{equation}
This can be generalized by replacing $2N$ with an arbitrary natural number, $\N$, and hence $\N$-extended Majorana  supersymmetry algebra. \newline

\indent $P_{\mu},\,Q^{a\alpha},\,1\leq a\leq\N$  generate a
supergroup, $\mbox{G}_{T}$, 
with parameters,  
$z^{M}= (x^{\mu},\theta^{a\alpha})$,  which
are coordinates on superspace.  The  general element of $\mbox{G}_{T}$
is written in terms of these coordinates as
\begin{equation}
g(z)=\displaystyle{e^{i(x{\cdot P}+\bar{Q}_{a}\theta^{a})}}\,.
\label{generalelement}
\end{equation}
Corresponding to eq.(\ref{Majorana}) $\theta^{a}$ also satisfies the Majorana condition
\begin{equation}
\bt_{a}=\theta^{a\dagger}\gamma^{0}=-\theta^{a}{}^{t}\e\,,
\end{equation}
so that 
\begin{equation}
\begin{array}{cc}
\bar{Q}_{a}\theta^{a}=\bt_{a}Q^{a}\,,~~~~&~~~~g(z)^{\dagger}=g(z)^{-1}=g(-z)\,.
\end{array}
\end{equation}

\indent The Baker-Campbell-Haussdorff formula  with the
supersymmetry algebra~(\ref{3Dsusy})  gives
\begin{equation}
g(z_{1})g(z_{2})=g(z_{3})\,,
\label{groupproperty1}
\end{equation}
where
\begin{equation}
\begin{array}{cc}
x^{\mu}_{3}=x^{\mu}_{1}+x^{\mu}_{2}+i\bt_{1a}\gamma^{\mu}\theta_{2}^{a}\,,
~~~~&~~~~
\theta^{a}_{3}=\theta_{1}^{a}+\theta_{2}^{a}\,.
\end{array}
\label{groupproperty2}
\end{equation}

Letting $z_{1}\rightarrow -z_{2}$ we may get the 
supertranslation  invariant  one forms,  $e^{M}=(e^{\mu},
{\rm d}\theta^{a\alpha})$, where 
\begin{equation}
e^{\mu}(z)={\rm d}x^{\mu}-i\bt_{a}\gamma^{\mu}{\rm d}\theta^{a}\,.
\label{infint}
\end{equation}
The exterior derivative,~${\rm d}$, on superspace  is 
defined as
\begin{equation}
{\rm d}\equiv {\rm d}z^{M}\frac{\partial~}{\partial z^{M}}
=e^{M}D_{M}=e^{\mu}\partial_{\mu}-{\rm d}\theta^{a\alpha}D_{a\alpha}\,,
\label{d}
\end{equation}
where $D_{M}=(\partial_{\mu},-D_{a\alpha})$ are 
covariant  derivatives  
\begin{equation}
\begin{array}{cc}
\displaystyle{\partial_{\mu}=\frac{\partial~}{\partial x^{\mu}}}\,,~~~~&~~~~
\displaystyle{D_{a\alpha}=-\frac{\partial~}{\partial\theta^{a\alpha}}+
i(\bt_{a}\gamma^{\mu})_{\alpha}\frac{\partial~}{\partial x^{\mu}}}\,.
\end{array}
\end{equation}
We also define
\begin{equation}
\bar{D}^{a\alpha}=\e^{-1\alpha\beta}D_{a\beta}
=\displaystyle{\frac{\partial~}{\partial\bt_{a\alpha}}-
i(\gamma^{\mu}\theta^{a})^{\alpha}\frac{\partial~}{\partial x^{\mu}}}\,,
\end{equation}
satisfying the anti-commutator relations
\begin{equation}
\displaystyle{
\{\bar{D}^{a\alpha},D_{b\beta}\}=
2i\delta^{a}{}_{b}\gamma^{\mu}{}^{\alpha}{}_{\beta}\partial_{\mu}}\,.
\label{anticomD}
\end{equation}
\newline
\indent Under an arbitrary superspace coordinate transformation,   
$z \longrightarrow z^{\prime}$,   
$e^{M}$ and $D_{M}$  transform as
\begin{equation}
\begin{array}{cc}
e^{M}(z^{\prime})
=e^{N}(z){\cal R}_{N}{}^{M}(z)\,,~~~~&~~~~
D^{\prime}_{M}={\cal R}^{-1}{}_{M}{}^{N}(z)D_{N}\,,
\end{array}
\label{ReD}
\end{equation}
so that the exterior derivative is left invariant
\begin{equation}
e^{M}(z)D_{M}=e^{M}(z^{\prime})D^{\prime}_{M}\,,
\end{equation}
where ${\cal R}_{M}{}^{N}(z)$ is a $(3+2{\cal N})\times (3+2{\cal N})$ 
supermatrix of the form 
\begin{equation}
{\cal R}_{M}{}^{N}(z)=
\left(\begin{array}{rr}
R_{\mu}{}^{\nu}(z)&\partial_{\mu}\theta^{\prime b\beta}\\
-B^{\mu}_{a\alpha}(z)&-D_{a\alpha}\theta^{\prime b\beta}
\end{array}\right)\,,
\label{calRorig}
\end{equation}
with
\begin{eqnarray}
&\displaystyle{R_{\mu}{}^{\nu}(z)=\frac{\partial 
x^{\prime \nu}}{\partial x^{\mu}}
-i\bt_{a}^{\prime}\gamma^{\nu}
\frac{\partial\theta^{\prime a}}{\partial 
x^{\mu}}}\,,~~~~~&\label{Rh}\\
&{}&\nonumber\\
&B^{\mu}_{a\alpha}(z)=D_{a\alpha}x^{\prime\mu}+
i\bt_{b}^{\prime}\gamma^{\mu}D_{a\alpha}\theta^{\prime b}\,.&\label{B}
\end{eqnarray}
\newline
\indent For Majorana spinors 
it is useful to note from eqs.(\ref{dagger},\ref{charge},\ref{L3}) 
\begin{subeqnarray}
&\bar{\varepsilon}_{a}\rho^{a}=\bar{\rho}_{a}\varepsilon^{a}\,,&\\
{}\nonumber\\
&\rho^{a}\bar{\varepsilon}_{a}+\varepsilon^{a}\bar{\rho}_{a}+\bar{\rho}_{a}\varepsilon^{a}\,1=0\,,&\\
{}\nonumber\\
&\bar{\rho}_{a}
\gamma^{\mu_{1}}\gamma^{\mu_{2}}\cdots\gamma^{\mu_{n}}\varepsilon^{a}
=(-1)^{n}\bar{\varepsilon}_{a}\gamma^{\mu_{n}}\cdots\gamma^{\mu_{2}}
\gamma^{\mu_{1}}\rho^{a}\,,&\\
{}\nonumber\\
&(\bar{\rho}_{a}
\gamma^{\mu_{1}}\gamma^{\mu_{2}}\cdots\gamma^{\mu_{n}}\varepsilon^{a})^{\ast}
=\bar{\varepsilon}_{a}\gamma^{\mu_{n}}\cdots\gamma^{\mu_{2}}
\gamma^{\mu_{1}}\rho^{a}\,.&
\end{subeqnarray}
In particular
\begin{equation}
\theta^{a}\bt_{a}=-\textstyle{\frac{1}{2}}\bt_{a}\theta^{a}\,1\,.
\label{thetatheta}
\end{equation}


\section{Superconformal Symmetry in Three-dimensions\label{scfs}}
In this section  we  first define the three-dimensional 
superconformal group on superspace and then discuss  its    
superconformal Killing equation  along with the solutions.
\subsection{Superconformal Group \&  Killing Equation}
The superconformal group is  defined here as a group of 
superspace coordinate
transformations, $z \stackrel{g}{\longrightarrow} z^{\prime}$,  that 
preserve the 
infinitesimal supersymmetric interval length, $e^{2}=\eta_{\mu\nu}
e^{\mu}e^{\nu}$, up to a local scale factor, so that
\begin{equation}
{e^{2}}(z)~\rightarrow~e^{2}(z^{\prime})=\Omega^{2}(z;g)e^{2}(z)\,, 
\label{scdef}
\end{equation}
where $\Omega(z;g)$ is a local scale factor.\newline
This requires $B^{\mu}_{a\alpha}(z)=0$ 
\begin{equation}
D_{a\alpha}x^{\prime\mu}+
i\bt_{b}^{\prime}\gamma^{\mu}D_{a\alpha}\theta^{\prime b}=0\,.
\label{B=0}
\end{equation}
and
\begin{eqnarray}
&e^{\mu}(z^{\prime})=e^{\nu}(z)R_{\nu}{}^{\mu}(z;g)\,,&\label{ehomogeneous}\\
&{}&\nonumber\\
&R^{~\lambda}_{\mu}(z;g)R^{~\rho}_{\nu}(z;g)\eta_{\lambda\rho}
=\Omega^{2}(z;g)\eta_{\mu\nu}\,,~~~~~~~~~
\det R(z;g)=\Omega^{3}(z;g)\,.\label{R2}&
\end{eqnarray}
Hence ${\cal R}_{M}{}^{N}$ in eq.(\ref{calRorig}) is of the 
form\footnote{More explicit form of ${\cal R}_{M}{}^{N}$ is obtained later 
in eq.(\ref{calRR}).}
\begin{equation}
{\cal R}_{M}{}^{N}(z;g)=
\left(\begin{array}{cr}
R^{~\nu}_{\mu}(z;g)&\partial_{\mu}\theta^{\prime b\beta}\\
0&-D_{a\alpha}\theta^{\prime b\beta}
\end{array}\right)\,.
\label{calR}
\end{equation}
\newline
\indent Infinitesimally $z^{\prime}\simeq z+\delta z$, 
eq.(\ref{B=0})  gives
\begin{equation}
D_{a\alpha}h^{\mu}=2i(\bar{\lambda}_{a}\gamma^{\mu})_{\alpha}\,,
\label{enough1}
\end{equation}
or equivalently
\begin{equation}
\bar{D}^{a\alpha}h^{\mu}=-2i(\gamma^{\mu}\lambda^{a})^{\alpha}\,,
\label{enough2}
\end{equation}
where we define
\begin{equation}
\begin{array}{cc}
\lambda^{a}=\delta\theta^{a}\,,~~~&~~~
\bar{\lambda}_{a}=\delta\bar{\theta}_{a}\,,\\
{}&{}\\
\multicolumn{2}{c}{
h^{\mu}=\delta x^{\mu}-i\bt_{a}\gamma^{\mu}\delta\theta^{a}\,.}
\end{array}
\end{equation}
Infinitesimally from eq.(\ref{Rh})  $R_{\mu}{}^{\nu}$ is of the form
\begin{equation} 
R_{\mu}{}^{\nu}\simeq\delta_{\mu}^{~\nu}+\partial_{\mu}h^{\nu}\,,
\label{infR}
\end{equation}
so that the condition~(\ref{R2}) reduces to the ordinary conformal
Killing equation
\begin{equation}
\partial_{\mu}h_{\nu}+\partial_{\nu}h_{\mu}\propto \eta_{\mu\nu}\,.
\label{ordiKi}
\end{equation}
We note that eq.(\ref{ordiKi}) follows from eq.(\ref{enough1}).  Using   
the anti-commutator relation for  $D_{a\alpha}$~(\ref{anticomD})  
we get from eqs.(\ref{enough1},\ref{enough2},\ref{L1})
\begin{equation}
\delta^{a}{}_{b}\partial_{\nu}h_{\mu}
=\textstyle{\frac{1}{2}}\left(
\bar{D}^{a\alpha}(\bar{\lambda}_{b}\gamma_{\mu}\gamma_{\nu})_{\alpha}
-(\gamma_{\nu}\gamma_{\mu}D_{b\alpha}\lambda^{a})^{\alpha}\right)\,,
\end{equation}
and hence
\begin{equation}
\delta_{a}^{~b}(\partial_{\mu}h_{\nu}+\partial_{\nu}h_{\mu})
=(\bar{D}^{a\alpha}\bar{\lambda}_{b\alpha}-D_{b\alpha}\lambda^{a\alpha})
\eta_{\mu\nu}\,,
\label{ordinaryKilling}
\end{equation}
which implies eq.(\ref{ordiKi}).   
Thus eq.(\ref{enough1}) or eq.(\ref{enough2}) 
is  a necessary and  sufficient  condition for a 
supercoordinate transformation to be superconformal.\newline

\indent From eq.(\ref{enough1},\ref{enough2}) $\lambda^{a\alpha},
\bar{\lambda}_{a\alpha}$ are given by
\begin{equation}
\begin{array}{cc}
\lambda^{a\alpha}=i\textstyle{\frac{1}{6}}\bar{D}^{a\beta}
{\rm h}^{\alpha}{}_{\beta}\,,~~~~&~~~~
\bar{\lambda}_{a\alpha}=-i\textstyle{\frac{1}{6}}D_{a\beta}
{\rm h}^{\beta}{}_{\alpha}\,,
\end{array}
\end{equation}
where 
\begin{equation}
{\rm h}^{\alpha}{}_{\beta}=h^{\mu}\gamma_{\mu}{}^{\alpha}{}_{\beta}\,.
\end{equation}
Substituting these expressions back into eqs.(\ref{enough1},\ref{enough2}) gives using eqs.(\ref{gg},\ref{complete})
\begin{equation}
\begin{array}{l}
D_{a\alpha}h^{\mu}=-i\textstyle{\frac{1}{2}}\e^{\mu}{}_{\nu\lambda}D_{a\beta}h^{\nu}\gamma^{\lambda}{}^{\beta}{}_{\alpha}\,,\\
{}\\
\bar{D}^{a\alpha}h^{\mu}=i\textstyle{\frac{1}{2}}\e^{\mu}{}_{\nu\lambda}
\gamma^{\lambda}{}^{\alpha}{}_{\beta}\bar{D}^{a\beta}h^{\nu}\,.
\end{array}
\label{Killing1}
\end{equation}
or equivalently
\begin{equation}
\begin{array}{l}
D_{a\alpha}\h^{\beta}{}_{\gamma}=\textstyle{\frac{2}{3}}\delta_{\alpha}{}^{\beta}D_{a\delta}\h^{\delta}{}_{\gamma}-\textstyle{\frac{1}{3}}\delta^{\beta}{}_{\gamma}D_{a\delta}\h^{\delta}{}_{\alpha}\,,\\
{}\\
\bar{D}^{a\alpha}\h^{\beta}{}_{\gamma}=\textstyle{\frac{2}{3}}\delta^{\alpha}{}_{\gamma}
\bar{D}^{a\delta}\h^{\beta}{}_{\delta}-\textstyle{\frac{1}{3}}\delta^{\beta}{}_{\gamma}\bar{D}^{a\delta}\h^{\alpha}{}_{\delta}\,.
\end{array}
\label{Killing2}
\end{equation}
Eq.(\ref{Killing1}) or eq.(\ref{Killing2}) 
may therefore be regarded as the fundamental
superconformal Killing equation and its solutions  give 
the generators of  extended superconformal transformations in 
three-dimensions. The general solution is\footnote{A method of obtaining 
the solution~(\ref{solutionforh}) is demonstrated  
in Appendix \ref{Appendixsolving}.} 
\begin{equation}
\begin{array}{ll}
h^{\mu}(z)=&2x{\cdot b}\,x^{\mu}-(x^{2}-
\textstyle{\frac{1}{4}}(\bt_{a}\theta^{a})^{2})b^{\mu}+
\e^{\mu}{}_{\nu\lambda}x^{\nu}b^{\lambda}\bt_{a}\theta^{a}\\
{}&{}\\
{}&\,-2\bar{\rho}_{a}\x_{-}\gamma^{\mu}\theta^{a}+
w^{\mu}{}_{\nu}x^{\nu}+\textstyle{\frac{1}{4}}
\epsilon^{\mu}{}_{\nu\lambda}w^{\nu\lambda}\bt_{a}\theta^{a}+\lambda x^{\mu}\\
{}&{}\\
{}&\,+it_{a}{}^{b}\bt_{b}\gamma^{\mu}\theta^{a}
+2i\bar{\varepsilon}_{a}\gamma^{\mu}\theta^{a}+a^{\mu}\,,
\end{array}
\label{solutionforh}
\end{equation}
where $a^{\mu},b^{\mu},\lambda,w_{\mu\nu}=-w_{\nu\mu}$ are real,\, $\varepsilon^{a},\rho^{a}$ satisfy the Majorana condition~(\ref{Majorana}) and\\ 
$t\in\mbox{so}(\N)$ satisfying
\begin{equation}
t^{\dagger}=t^{t}=-t\,.
\label{Rcondition}
\end{equation}
We also set 
\begin{equation}
\begin{array}{cc}
\x=x^{\mu}\gamma\,,~~~~&~~~~\x_{\pm}=\x\pm i\textstyle{\frac{1}{2}}\bt_{a}\theta^{a}\,1\,.
\end{array}
\end{equation}
Eq.(\ref{solutionforh}) gives
\begin{equation}
\lambda^{a}=\x_{+}b{\cdot\gamma}\theta^{a}-i\x_{+}\rho^{a}
+2(\bar{\rho}_{b}\theta^{a})\theta^{b}
+(w+\textstyle{\frac{1}{2}}\lambda)\theta^{a}-\theta^{b}t_{b}{}^{a}
+\varepsilon^{a}
\label{sol1}
\end{equation}
satisfying the  Majorana condition
\begin{equation}
\bar{\lambda}_{a}=\lambda^{a}{}^{\dagger}\gamma^{0}=-\lambda^{at}\e\,,
\end{equation}
where we put
\begin{equation}
w=\textstyle{\frac{1}{4}}w_{\mu\nu}\gamma^{\mu}\gamma^{\nu}\,.
\label{wdef}
\end{equation}
For later use it is worth to note
\begin{equation}
\begin{array}{cc}
\gamma^{0}w\gamma^{0}=-w^{\dagger}\,,~~~~&~~~~\e w\e^{-1}=-w^{t}\,.
\end{array}
\label{wt}
\end{equation}

\subsection{Extended Superconformal Transformations} 
In summary, the generators of  superconformal transformations in
three-dimensions acting  on the  three-dimensional  superspace, 
$\Real^{3|2{\cal N}}$, with coordinates, $z^{M}
=(x^{\mu},\theta^{a\alpha})$,  can be classified as  
\begin{enumerate}
\item  Supertranslations, $a,\varepsilon$
\begin{equation}
\begin{array}{cc}
\delta x^{\mu}=a^{\mu}-i\bt_{a}\gamma^{\mu}\varepsilon^{a}\,,~~~~&~~~~
\delta\theta^{a}=\varepsilon^{a}\,.
\end{array}
\label{trans}
\end{equation}
This is consistent with  eq.(\ref{groupproperty2}).
\item Dilations, $\lambda$
\begin{equation}
\begin{array}{cc}
\delta x^{\mu}=\lambda x^{\mu}\,,~~~~&~~~~
\delta\theta^{a}=\textstyle{\frac{1}{2}}\lambda\theta^{a}\,.
\end{array}
\end{equation}
\item Lorentz transformations, $w$ 
\begin{equation}
\begin{array}{cc}
\delta x^{\mu}=w^{\mu}{}_{\nu}x^{\nu}\,,~~~~&~~~~ 
\delta\theta^{a}=w\theta^{a}\,.
\end{array}
\end{equation}
\item $R$-symmetry transformations, $t$
\begin{equation}
\begin{array}{cc}
\delta x^{\mu}=0\,,~~~~&~~~~
\delta\theta^{a}=-\theta^{b}t_{b}{}^{a}\,.
\end{array}
\end{equation}
where $t\in\mbox{so}(\N)$ of dimension $\frac{1}{2}\N(\N-1)$. 
\item Special superconformal transformations, $b,\rho$ 
\begin{equation}
\begin{array}{l}
\delta x^{\mu}=2x{\cdot b}\,x^{\mu}-(x^{2}+\textstyle{\frac{1}{4}}(\bt_{a}\theta^{a})^{2})b^{\mu}
-\bar{\rho}_{a}\x_{+}\gamma^{\mu}\theta^{a}\,,\\
{}\\
\delta\theta^{a}=\x_{+}b{\cdot\gamma}\theta^{a}-i\x_{+}\rho^{a}+2(\bar{\rho}_{b}\theta^{a})\theta^{b}\,.
\end{array}
\label{special}
\end{equation}
\end{enumerate}

\indent As we consider infinitesimal transformations we obtain $\mbox{SO}(\N)$ as $R$-symmetry group.    
However finitely $R$-symmetry group can be extended  to $\mbox{O}(\N)$  
which  leaves the supertranslation invariant one form~(\ref{infint})   invariant manifestly.    
\newpage
\subsection{Superinversion}
In three-dimensions we define superinversion, 
$z^{M}\stackrel{i_{s}}{\longrightarrow}z^{\prime M}
=(x^{\prime\mu},\theta^{\prime a\alpha})\in\Real^{3|2{\cal N}}$, by
\begin{equation}
\begin{array}{cc}
\x^{\prime}_{\pm}=-\x_{\pm}^{-1}\,,~~~~&~~~~
\theta^{\prime}{}^{a}=i\x_{+}^{-1}\theta^{a}\,.
\end{array}
\label{superinversion}
\end{equation}
As a consistency check we note from $\x_{+}\x_{-}=(x^{2}+\textstyle{\frac{1}{4}}(\bt_{a}\theta^{a})^{2})\,1$
\begin{equation}
\begin{array}{cc}
\bt^{\prime}_{a}=\theta^{\prime a\dagger}\gamma^{0}=
-\theta^{\prime a}{}^{t}\e\,,~~~~&~~~~
\x_{+}^{\prime}-\x^{\prime}_{-}=i\bt^{\prime}_{a}\theta^{\prime a}\,1\,.
\end{array}
\end{equation}
It is easy to verify that superinversion is idempotent
\begin{equation}
i_{s}^{2}=1\,.
\end{equation}
Using
\begin{equation}
{\rm e}(z)=e^{\mu}(z)\gamma_{\mu}
={\rm d}{\rm x}_{+}+2i{\rm d}\theta^{a}\bt_{a}\,,
\end{equation}
we get under superinversion
\begin{equation}
{\rm e}(z^{\prime})={\rm x}_{+}^{-1}{\rm e}(z){\rm x}_{-}^{-1}\,.
\label{ee}
\end{equation}
and hence
\begin{equation}
\begin{array}{cc}
e^{2}(z^{\prime})=\Omega^{2}(z;i_{s})e^{2}(z)\,,~~~~&~~~~
\Omega(z;i_{s})=\displaystyle{\frac{1}{x^{2}+\frac{1}{4}(\bt_{a}\theta^{a})^{2}}}\,.
\end{array}
\label{Oi}
\end{equation}
Eq.(\ref{ee}) can be rewritten as 
\begin{equation}
\begin{array}{cc}
e^{\mu}(z^{\prime})=e^{\nu}(z)R_{\nu}{}^{\mu}(z;i_{s})\,,~~~~&~~~~
R_{\nu}{}^{\mu}(z;i_{s})=\textstyle{\frac{1}{2}}
\mbox{tr}(\gamma_{\nu}\x_{-}^{-1}\gamma^{\mu}\x_{+}^{-1})\,.
\end{array}
\label{Rzi}
\end{equation}
Explicitly
\begin{equation}
R_{\nu}{}^{\mu}(z;i_{s})
=\displaystyle{\frac{1}{(x^{2}+\frac{1}{4}(\bt_{a}\theta^{a})^{2})^{2}}}
\left(2x_{\nu}x^{\mu}-(x^{2}-\textstyle{\frac{1}{4}}(\bt_{a}\theta^{a})^{2})\delta_{\nu}{}^{\mu}-\e_{\nu}{}^{\mu\lambda}x_{\lambda}\bt_{a}\theta^{a}\right)\,.
\end{equation}
Note that
\begin{equation}
\begin{array}{cc}
\gamma^{\nu}R_{\nu}{}^{\mu}(z;i_{s})=\x^{-1}_{-}\gamma^{\mu}\x_{+}^{-1}\,,
~~~~&~~~~R_{\nu}{}^{\mu}(z;i_{s})\gamma_{\mu}=\x^{-1}_{+}\gamma_{\nu}\x_{-}^{-1}\,.
\end{array}
\label{Rgamma}
\end{equation}

\indent If we consider a transformation,$~z\,
\stackrel{i_{s}\circ g\circ i_{s}}{-\!\!\!-\!\!\!-\!\!\!\longrightarrow}\,z^{\prime}$, where 
$g$ is a three-dimensional superconformal transformation, then  
we get 
\begin{equation}
\begin{array}{ll}
h^{\mu}(z)=&2x{\cdot a}\,x^{\mu}-(x^{2}-
\textstyle{\frac{1}{4}}(\bt_{a}\theta^{a})^{2})a^{\mu}+
\e^{\mu}{}_{\nu\lambda}x^{\nu}a^{\lambda}
\bt_{a}\theta^{a}\\
{}&{}\\
{}&\,-2\bar{\varepsilon}_{a}\x_{-}\gamma^{\mu}\theta^{a}+
w^{\mu}{}_{\nu}x^{\nu}+\textstyle{\frac{1}{4}}
\epsilon^{\mu}{}_{\nu\lambda}w^{\nu\lambda}\bt_{a}\theta^{a}
-\lambda x^{\mu}\\
{}&{}\\
{}&\,+it_{a}{}^{b}\bt_{b}\gamma^{\mu}\theta^{a}
+2i\bar{\rho}_{a}\gamma^{\mu}\theta^{a}+b^{\mu}\,.
\end{array}
\end{equation}
Hence, under  superinversion, the superconformal transformations
are related by 
\begin{equation}
\begin{array}{ccc}
{\cal K}\equiv\left(\begin{array}{c}
a^{\mu}\\
b^{\mu}\\
\varepsilon^{a}\\
\rho^{a}\\
\lambda\\
w^{\mu}{}_{\nu}\\
t^{a}{}_{b}
\end{array}\right)~~~~& \longrightarrow &~~~~
\left(\begin{array}{c}
b^{\mu}\\
a^{\mu}\\
\rho^{a}\\
\varepsilon^{a}\\
-\lambda\\
w^{\mu}{}_{\nu}\\
t^{a}{}_{b}
\end{array}\right)\,.
\end{array}
\end{equation}
In particular, special superconformal transformations~(\ref{special})
can be obtained by 
\begin{equation}
z\,\stackrel{i_{s}\circ (b,\rho)\circ
i_{s}}{-\!\!\!-\!\!\!-\!\!\!-\!\!\!-\!\!\!-\!\!\!\longrightarrow}\,z^{\prime}
\,, 
\label{specialz}
\end{equation}
where $(b,\rho)$ is a supertranslation.

\subsection{Superconformal Algebra\label{SCA}}
The generator of  infinitesimal  
superconformal transformations,~${\cal L}$,
is given by
\begin{equation}
{\cal L}
=h^{\mu}\partial_{\mu}-\lambda^{a\alpha}D_{a\alpha}\,.
\label{calL}
\end{equation}
If we write the commutator of two generators,~${\cal L}_{1},{\cal
L}_{2}$,  as
\begin{equation}
[{\cal L}_{2},{\cal L}_{1}]={\cal L}_{3}
=h^{\mu}_{3}\partial_{\mu}-\lambda^{a\alpha}_{3}D_{a\alpha}\,,
\label{hcommutator}
\end{equation}
then $h_{3}^{\mu},\,\lambda^{a\alpha}_{3}$ 
are given by 
\begin{equation}
\begin{array}{l}
h^{\mu}_{3}=h^{\nu}_{2}\partial_{\nu}h^{\mu}_{1}-h^{\nu}_{1}\partial_{\nu}h^{\mu}_{2}+2i\bar{\lambda}_{1a}\gamma^{\mu}\lambda^{a}_{2}\,,\\
{}\\
\lambda^{a}_{3}={\cal L}_{2}\lambda_{1}^{a}-{\cal L}_{1}\lambda_{2}^{a}\,,
\end{array}
\end{equation}
and $h^{\mu}_{3},\,\lambda^{a}_{3}$ 
satisfy eq.(\ref{enough2}) verifying the closure of the Lie algebra.\newline
Explicitly with eqs.(\ref{solutionforh}\,,\ref{sol1}) 
we get
\begin{equation}
\begin{array}{l}
\displaystyle{a^{\mu}_{3}=w^{\mu}_{1}{}_{\nu}a^{\nu}_{2}
+\lambda_{1}a^{\mu}_{2}+
i\bar{\varepsilon}_{1a}\gamma^{\mu}\varepsilon_{2}^{a} -
(1\leftrightarrow 2)}\,,\\
{}\\
\varepsilon^{a}_{3}=
w_{1}\varepsilon_{2}^{a}+
\textstyle{\frac{1}{2}}\lambda_{1}\varepsilon^{a}_{2}-
ia_{2}{\cdot\gamma}\rho_{1}^{a}
-\varepsilon_{2}^{b}t_{1b}{}^{a}-(1\leftrightarrow 2)\,,\\
{}\\
\displaystyle{\lambda_{3}=2a_{2}{\cdot b_{1}}-2
\bar{\rho}_{1a}\varepsilon_{2}^{a}-(1\leftrightarrow 2)}\,,\\
{}\\
\displaystyle{w^{\mu\nu}_{3}=w^{\mu}_{1}{}_{\lambda}w^{\lambda\nu}_{2}
+2(a_{2}^{\mu}b_{1}^{\nu}-a_{2}^{\nu}b_{1}^{\mu})+
2\bar{\rho}_{1a}\gamma^{[\mu}\gamma^{\nu]}\varepsilon^{a}_{2}
-(1\leftrightarrow 2)}\,,\\
{}\\
b^{\mu}_{3}=w^{\mu}_{1}{}_{\nu}b^{\nu}_{2}-\lambda_{1}b^{\mu}_{2}+
i\bar{\rho}_{1a}\gamma^{\mu}\rho_{2}^{a} -(1\leftrightarrow 2)\,,\\
{}\\
\rho^{a}_{3}=w_{1}\rho_{2}^{a}-\textstyle{\frac{1}{2}}
\lambda_{1}\rho_{2}^{a}-ib_{2}{\cdot\gamma}\varepsilon_{1}^{a}-
\rho_{2}^{b}t_{1b}{}^{a}-(1\leftrightarrow 2)\,,\\
{}\\
\displaystyle{t_{3a}{}^{b}=(t_{1}t_{2})_{a}{}^{b}
+2(\bar{\rho}_{2a}\varepsilon^{b}_{1}-\bar{\varepsilon}_{1a}\rho_{2}^{b})
-(1\leftrightarrow 2)}\,.
\end{array}
\label{MMcom}
\end{equation}
From eq.(\ref{MMcom}) we can read off the explicit forms of 
three-dimensional  superconformal algebra as exhibited 
in Appendix \ref{AppendixC}.\newline

\indent If we define a $(4+2{\cal N})\times (4+2{\cal N})$ 
supermatrix,~$M$, as 
\begin{equation}
\displaystyle{
M=\left(\begin{array}{ccc}
w+\textstyle{\frac{1}{2}}\lambda&ia{\cdot\gamma}&\sqrt{2}\varepsilon^{b}\\
ib{\cdot\gamma}&w-{\textstyle\frac{1}{2}}\lambda&\sqrt{2}\rho^{b}\\
-\sqrt{2}\bar{\rho}_{a}&-\sqrt{2}\bar{\varepsilon}_{a}&t_{a}{}^{b}
\end{array}\right)}\,,
\label{Mform}
\end{equation}
then the relation above~(\ref{MMcom}) agrees with the matrix commutator  
\begin{equation}
[M_{1},M_{2}]=M_{3}\,.
\label{Mcommutator}
\end{equation}
In general, $M$ can be defined as a $(4,2{\cal N})$ supermatrix subject to 
\begin{subeqnarray}
\label{Mcondition}
&BMB^{-1}=-M^{\dagger}\,,~~~~~~~~
B=\left(\begin{array}{ccc}
                 0&\gamma^{0}&0\\ 
                 \gamma^{0}&0&0\\
                 0&0&1
              \end{array}\right)\,,&\label{Mcondition1}\\
\nonumber\\
&CMC^{-1}=-M^{t}\,,~~~~~~~~~~
C=\left(\begin{array}{ccc}
                 0&\e&0\\ 
                 \e&0&0\\
                 0&0&1
              \end{array}\right)\,.~&\label{Mcondition2}
\end{subeqnarray}
Supermatrix of the form~(\ref{Mform}) is the general solution of 
these two equations. \newline

\indent The $4\times 4$ matrix appearing in $M$,
\begin{equation}
\left(\begin{array}{cc}
w+\textstyle{\frac{1}{2}}\lambda &ia{\cdot\gamma}\\
ib{\cdot\gamma}&w-\textstyle{\frac{1}{2}}\lambda
\end{array}\right)\,,
\end{equation}
corresponds to a generator of $\mbox{SO}(2,3)\cong\mbox{Sp}(2,\Real)$ 
as demonstrated in Appendix \ref{AppendixD}. Thus,  
the $\N$-extended Majorana       
superconformal group in three-dimensions may be identified  
with the supermatrix group generated by supermatrices of the 
form  $M$~(\ref{Mform}), which is    
$\mbox{OSp}(\N|2,\Real)\equiv\mbox{G}_{S}$ having  dimensions     
$(10+\frac{1}{2}\N(\N-1)|4{\cal N})$.

\newpage

\section{Coset Realization of  Transformations\label{coset}}
In this section, we first obtain an explicit formula for  
the finite non-linear 
superconformal transformations of the supercoordinates and discuss several representations of the superconformal group. 
We then 
construct  matrix or vector valued 
functions depending on  two or   three points 
in superspace which transform 
covariantly under superconformal transformations.  
These  variables serve as  building blocks of    
obtaining two-point, three-point and general $n$-point 
correlation functions later.
\subsection{Superspace as a Coset}
To obtain an explicit  formula for the finite non-linear superconformal
transformations, we first identify the superspace, $\Real^{3|2{\cal N}}$,  
as a coset, 
$\mbox{G}_{S}/\mbox{G}_{0}$,
where $\mbox{G}_{0}\subset \mbox{G}_{S}$ is the subgroup generated by
matrices, $M_{0}$, 
of the form~(\ref{Mform}) with $a^{\mu}=0,\,\varepsilon^{a}=0$ and
depending on 
parameters $b^{\mu},\,\rho^{a},\,\lambda,\,w_{\mu\nu},\,t_{a}{}^{b}$.    
The group of supertranslations, $\mbox{G}_{T}$,  
parameterized by coordinates, 
$z^{M}\in\Real^{3|2{\cal N}}$,   has been defined by general
elements as in eq.(\ref{generalelement})   with the group property
given by eqs.(\ref{groupproperty1},\,\ref{groupproperty2}). Now we may
represent it by  
supermatrices\footnote{The subscript, $T$, denotes supertranslations.}
\begin{equation}
G_{T}(z)=\mbox{exp}\left(\begin{array}{ccc}
               0&i\x&\sqrt{2}\theta^{b}\\
               0&0&0\\
               0&-\sqrt{2}\bt_{a}&0
                \end{array}\right)=
\left(\begin{array}{ccc}
      1&  i\x_{-}&\sqrt{2}\theta^{b}\\
      0&  1&0\\
      0&  -\sqrt{2}\bt_{a}&\delta_{a}{}^{b}
        \end{array}\right)\,.
\end{equation}
Note   $G_{T}(z)^{-1}=G_{T}(-z)$.\newline

\indent  
In general an element of $\mbox{G}_{S}$  can be uniquely decomposed  as
$G_{T}G_{0}^{-1}$. Thus for any  element $G(g)\in\mbox{G}_{S}$ we may define a
superconformal transformation, $z \stackrel{g}{\longrightarrow}
z^{\prime}$, and an associated element $G_{0}(z;g)\in\mbox{G}_{0}$ by
\begin{equation}
G(g)^{-1}G_{T}(z)G_{0}(z;g)=G_{T}(z^{\prime})\,.
\label{GTFINITE}
\end{equation}
If $G(g)\in\mbox{G}_{T}$ then clearly $G_{0}(z;g)=1$.  
Infinitesimally eq.(\ref{GTFINITE}) becomes
\begin{equation}
\delta G_{T}(z)=MG_{T}(z)-G_{T}(z)\hat{M}_{0}(z)\,, 
\label{MGL}
\end{equation}
where  $M$ is given by eq.(\ref{Mform}) and 
$\hat{M}_{0}(z)$, the generator of $\mbox{G}_{0}$,  has the form 
\begin{equation}
\hat{M}_{0}(z)
=\displaystyle{\left(\begin{array}{ccc}
        \hat{w}(z) +\textstyle{\frac{1}{2}}\hat{\lambda}(z)&0&0\\
     ib{\cdot\gamma}&\hat{w}(z)-\frac{1}{2}\hat{\lambda}(z)
     &\sqrt{2}\hat{\rho}{}^{b}(z)\\
       -\sqrt{2}\hat{\bar{\rho}}_{a}(z)&0&\hat{t}{}_{a}{}^{b}(z)
        \end{array}\right)}\,.
\end{equation}                           
The components depending on $z$ are given by 
\begin{equation}
\begin{array}{l}
\hat{w}(z)+\textstyle{\frac{1}{2}}\hat{\lambda}(z)
=w+\textstyle{\frac{1}{2}}\lambda+\x_{+}b{\cdot\gamma}+2\theta^{a}\bar{\rho}_{a}\,,\\
{}\\
\hat{w}(z)-\textstyle{\frac{1}{2}}\hat{\lambda}(z)
=w-\textstyle{\frac{1}{2}}\lambda-b{\cdot\gamma}\x_{-}-2\rho^{a}\bt_{a}\,,\\
{}\\
\hat{\lambda}(z)=\lambda+2x{\cdot b}-2\bt_{a}\rho^{a}\,,\\
{}\\
\hat{\rho}^{a}(z)=\rho^{a}+ib{\cdot\gamma}\theta^{a}\,,\\
{}\\
\hat{\bar{\rho}}_{a}(z)=\bar{\rho}_{a}-i\bt_{a}b{\cdot\gamma}
=(\hat{\rho}^{a}(z))^{\dagger}\gamma^{0}=-\hat{\rho}^{a}(z)^{t}\e\,,\\
{}\\
\hat{t}{}_{a}{}^{b}(z)=t_{a}{}^{b}+2i\bt_{a}b{\cdot\gamma}\theta^{b}
+2\bt_{a}\rho^{b}-2\bar{\rho}_{a}\theta^{b}\,.
\end{array}
\label{hats}
\end{equation}
$\hat{w}(z)$ can be also written as 
$\hat{w}(z)=\textstyle{\frac{1}{4}}
\hat{w}_{\mu\nu}(z)\gamma^{\mu}\gamma^{\nu}$ with
\begin{equation}
\hat{w}_{\mu\nu}(z)=w_{\mu\nu}+2(x_{\mu}b_{\nu}-x_{\nu}b_{\mu})
+\epsilon_{\mu\nu\lambda}(b^{\lambda}\bt_{a}\theta^{a}+
2i\bar{\rho}_{a}\gamma^{\lambda}\theta^{a})\,.
\end{equation}
Writing $\delta G_{T}(z)={\cal L}G_{T}(z)$ we may verify that ${\cal
L}$ is identical to eq.(\ref{calL}).\newline

\indent The definitions\,(\ref{hats}) can be summarized by 
\begin{subeqnarray}
&D_{a\alpha}\lambda^{b\beta}(z)=
-\textstyle{\frac{1}{2}}\delta_{a}{}^{b}\delta_{\alpha}{}^{\beta}
\hat{\lambda}(z)-\delta_{a}{}^{b}\hat{w}^{\beta}{}_{\alpha}(z)+
\delta_{\alpha}{}^{\beta}\hat{t}{}_{a}{}^{b}(z)\,,&\\
{}\nonumber\\
&\partial_{\nu}h_{\mu}(z)=\hat{w}_{\mu\nu}(z)+\eta_{\mu\nu}\hat{\lambda}(z)
\,,&
\end{subeqnarray}
and they  give
\begin{equation}
[D_{a\alpha},{\cal L}]=-D_{a\alpha}\lambda^{b\beta}D_{b\beta}=
(\textstyle{\frac{1}{2}}\delta_{a}{}^{b}\delta_{\alpha}{}^{\beta}
\hat{\lambda}(z)+\delta_{a}{}^{b}\hat{w}^{\beta}{}_{\alpha}(z)-
\delta_{\alpha}{}^{\beta}\hat{t}{}_{a}{}^{b}(z))D_{b\beta}\,.
\label{comDL}
\end{equation}
For later use we note 
\begin{equation}
\begin{array}{l}
D_{a\alpha}\hat{w}_{\mu\nu}(z)=
2(\hat{\bar{\rho}}_{a}(z)\gamma_{[\mu}\gamma_{\nu ]})_{\alpha}\,,\\
{}\\
D_{a\alpha}\hat{\lambda}(z)=2\hat{\bar{\rho}}_{a\alpha}(z)\,,\\
{}\\
D_{a\alpha}\hat{t}{}_{b}{}^{c}(z)=2(\delta_{ab}\delta^{cd}-\delta_{a}{}^{c}\delta_{b}{}^{d})\hat{\bar{\rho}}_{d\alpha}\,.
\end{array}
\label{Ddelta}
\end{equation}
\newline

\indent  The above analysis  can be simplified by reducing $G_{0}(z;g)$. To
achieve this we let 
\begin{equation}
Z_{0}=\left(\begin{array}{cc}
              0&0\\
              1&0\\
              0&1
            \end{array}\right)\,,
\end{equation}
and then
\begin{equation}
\begin{array}{cc}
M_{0}Z_{0}=Z_{0}H_{0}\,,~~~~&~~~~
H_{0}=
\left(\begin{array}{cc}
w-\textstyle{\frac{1}{2}}\lambda&\sqrt{2}\rho^{b}\\
                             0&t_{a}{}^{b}
                          \end{array}\right)\,.
\end{array}
\label{M0Z0}
\end{equation}
Now if we define
\begin{equation}
Z(z)\equiv G_{T}(z)Z_{0}=
\left(\begin{array}{cc}
        i\x_{-}&\sqrt{2}\theta^{b}\\
        1&0\\
        -\sqrt{2}\bt_{a}&\delta_{a}{}^{b}
        \end{array}\right)\,,
\label{defZ}
\end{equation}
then $Z(z)$ transforms  
under infinitesimal  superconformal transformations as
\begin{equation}
\delta Z(z)={\cal L}Z(z)=MZ(z)-Z(z)H(z)\,, 
\label{infZ}
\end{equation}
where $H(z)$ is given by  
\begin{equation}
\begin{array}{ll}
\hat{M}_{0}(z)Z_{0}=Z_{0}H(z)\,,&
H(z)=
\left(\begin{array}{cc}
\hat{w}(z)-\textstyle{\frac{1}{2}}\hat{\lambda}(z)&\sqrt{2}\hat{\rho}^{b}(z)\\
                             0&\hat{t}{}_{a}{}^{b}(z)
                          \end{array}\right)\,.
\end{array}
\label{defH}
\end{equation}

\indent  From eqs.(\ref{hcommutator},\,\ref{Mcommutator}) considering 
\begin{equation}
[{\cal L}_{2},{\cal L}_{1}]Z(z)={\cal L}_{3}Z(z)\,,
\label{defL3}
\end{equation}
we get
\begin{equation}
H_{3}(z)={\cal L}_{2}H_{1}(z)-{\cal L}_{1}H_{2}(z)+[H_{1}(z),H_{2}(z)]\,,
\label{Kcom}
\end{equation}
which gives separate equations for
$\hat{w},\,\hat{\lambda},\,\hat{\rho}$ and $\hat{t}{}_{a}{}^{b}$, thus 
$\hat{\lambda}_{3}={\cal L}_{2}\hat{\lambda}_{1}-{\cal 
L}_{1}\hat{\lambda}_{2}$, etc.\newline

\indent    As a conjugate of  $Z(z)$ we   define  $\bar{Z}(z)$ by
\begin{equation}
\bar{Z}(z)=\left(\begin{array}{cc}
               \gamma^{0}&0\\
                0&1\end{array}\right)Z(z){}^{\dagger}B
          =\left(\begin{array}{cc}
               \e^{-1}&0\\
               0&1\end{array}\right)Z(z){}^{t}C
          =\left(\begin{array}{ccc}
                 1&-i\x_{+}&-\sqrt{2}\theta^{b}\\
                 0&\sqrt{2}\bt_{a}&\delta_{a}{}^{b}\end{array}\right)\,.
\label{defbarZ}
\end{equation}
This satisfies
\begin{equation}
\bar{Z}(z)=\bar{Z}(0)G_{T}(z)^{-1}\,,
\end{equation}
and corresponding to eq.(\ref{infZ}) we have
\begin{equation}
\delta \bar{Z}(z)={\cal L}\bar{Z}(z)
=\bar{H}(z)\bar{Z}(z)-\bar{Z}(z)M \,,
\label{infbarZ}
\end{equation}
where
\begin{equation}
\bar{H}(z)=\left(\begin{array}{cc}
\hat{w}(z)+\textstyle{\frac{1}{2}}\hat{\lambda}(z)&0\\
-\sqrt{2}\hat{\bar{\rho}}{}_{a}(z)&\hat{t}{}_{a}{}^{b}(z)
\end{array}\right)\,.
\end{equation}

\subsection{Finite Transformations}
Finite superconformal transformations can be  obtained by exponentiation  
of infinitesimal transformations. To obtain a superconformal
transformation, $z\stackrel{g}{\longrightarrow}z^{\prime}$, we
therefore solve the differential equation
\begin{equation}
\begin{array}{ccc}
\displaystyle{
\frac{{\rm d}~}{{\rm d}t}z^{M}_{t}={\cal L}^{M}(z_{t})}\,,~~~~&~~~~z_{0}=z\,,
~~~&~~~z_{1}=z^{\prime}\,,
\end{array}
\label{zt01}
\end{equation}
where, with ${\cal L}$ given in eq.(\ref{calL}), ${\cal L}^{M}(z)$ is defined by 
\begin{equation}
{\cal L}={\cal L}^{M}(z)\partial_{M}\,.
\end{equation}
From eq.(\ref{infZ}) we get
\begin{equation}
\displaystyle{
\frac{{\rm d}~}{{\rm d}t}Z(z_{t})=MZ(z_{t})-Z(z_{t})H(z_{t}) }\,,
\end{equation}
which integrates to
\begin{equation}
Z(z_{t})=e^{tM}Z(z)K(z,t)\,,
\label{finites}
\end{equation}
where $K(z,t)$ satisfies
\begin{equation}
\begin{array}{cc}
\displaystyle{
\frac{{\rm d}~}{{\rm d}t}K(z,t)=-K(z,t)H(z_{t})}\,,~~~~&~~~~
K(z,0)=\left(\begin{array}{cc}
               1&0\\
               0&1
             \end{array}\right)\,.
\end{array}
\end{equation}
Hence for $t=1$ with $K(z,1)\equiv K(z;g)$,  $z\stackrel{g}{\longrightarrow}z^{\prime}$, eq.(\ref{finites}) becomes
\begin{equation}
\begin{array}{cc}
Z(z^{\prime})=G(g)^{-1}Z(z)K(z;g)\,,~~~~&~~~~G(g)^{-1}=e^{M}\,.
\end{array}
\label{Ztr}
\end{equation}
$G_{0}(z;g)$ in eq.(\ref{GTFINITE}) is related to 
$K(z;g)$ from eq.(\ref{Ztr})  by
\begin{equation}
G_{0}(z;g)Z_{0}=Z_{0}K(z;g)\,.
\end{equation}
In general  $K(z;g)$ is of the form
\begin{equation}
K(z;g)=\left(\begin{array}{cc}
                \Omega(z;g)^{\frac{1}{2}}L(z;g)&\sqrt{2}\Sigma^{b}(z;g)\\
                 0&U_{a}{}^{b}(z;g)
                \end{array}\right)\,,
\label{Kform}
\end{equation}
where $\Omega(z;g)$  is identical to the local scale factor in 
eq.(\ref{scdef}),  $U(z;g)\in\mbox{SO}(\N)$     
\begin{equation}
\begin{array}{cc}
U^{-1}=U^{\dagger}=U^{t}\,,~~~~&~~~~\det U=1\,,  
\end{array}
\end{equation}
and $L(z;g)$ satisfies
\begin{subeqnarray}
&\det L(z;g)=1\,,&\\
{}\nonumber\\
&L^{-1}(z;g)=\e^{-1}L(z;g)^{t}\e=\gamma^{0}L(z;g)^{\dagger}\gamma^{0}\,.&
\end{subeqnarray}

\indent From eq.(\ref{Ztr}) $\bar{Z}(z)$ transforms as
\begin{equation}
\bar{Z}(z^{\prime})=\bar{K}(z;g)\bar{Z}(z)G(g)\,,
\label{barZtr}
\end{equation}
where
\begin{equation}
\begin{array}{ll}
\bar{K}(z;g)&=\left(\begin{array}{cc}
               \gamma^{0}&0\\
                0&1\end{array}\right)K(z){}^{\dagger}\left(\begin{array}{cc}
               \gamma^{0}&0\\
                0&1\end{array}\right)
          =\left(\begin{array}{cc}
               \e^{-1}&0\\
               0&1\end{array}\right)K(z)^{t}\left(\begin{array}{cc}
               \e&0\\
               0&1\end{array}\right)\\
{}&{}\\
{}&=\left(\begin{array}{cc}
 \Omega(z;g)^{\frac{1}{2}}L^{-1}(z;g)&0\\
   \sqrt{2}\bar{\Sigma}_{a}(z;g)&U^{-1}{}_{a}{}^{b}(z;g)
                \end{array}\right)\,.
\end{array}
\label{Kbarform}
\end{equation}

\indent    If we define for superinversion, 
$z\stackrel{i_{s}}{\longrightarrow}z^{\prime}$,~(\ref{superinversion})  
\begin{equation}
\begin{array}{cc}
G(i_{s})^{-1}=\left(\begin{array}{ccc}
\epsilon&0&0\\
0&\epsilon&0\\
0&0&1
\end{array}\right)\,,~~~&~~~
K(z;i_{s})=\left(\begin{array}{cc}
-i(\e\x_{-})^{-1}&\sqrt{2}i\x^{-1}_{+}\theta^{b}\\
0&V_{a}{}^{b}(z)
\end{array}\right)\,,
\end{array}
\end{equation}
with
\begin{equation}
V_{a}{}^{b}(z)=\delta_{a}{}^{b}+2i\bt_{a}\x_{+}^{-1}\theta^{b}\,,
\label{Vform}
\end{equation}
an analogous formula  to eq.(\ref{Ztr}) can be obtained  for superinversion
\begin{equation}
G(i_{s})^{-1}Z(z)K(z;i_{s})
=\left(\begin{array}{cc}
1&0\\
-i\x^{\prime t}_{+}&\sqrt{2}\bt_{a}^{\prime t}\\
\sqrt{2}\theta^{\prime bt}& \delta^{b}{}_{a}
\end{array}\right)=
\bar{Z}(z^{\prime})^{t}\,.
\label{tZp1}
\end{equation}
Similarly we have
\begin{equation}
\bar{K}(z;i_{s})\bar{Z}(z)G(i_{s})=Z(z^{\prime})^{t}\,,
\label{tZp2}
\end{equation}
where
\begin{equation}
\bar{K}(z;i_{s})=\left(\begin{array}{cc}
i\e\x_{+}^{-1}&0\\
-\sqrt{2}i\bt_{a}\x^{-1}_{-}&V^{-1}{}_{a}{}^{b}(z)
\end{array}\right)\,.
\end{equation}
Note that
\begin{subeqnarray}
&V^{-1}(z)=V{}^{\dagger}=V(z)^{t}=V(-z)=1-2i\bt\x_{-}^{-1}\theta\,,&\label{V-z}\\
{}\nonumber\\
&\theta^{a}V_{a}{}^{b}(z)=\x_{-}\x_{+}^{-1}\theta^{b}\,,~~~~~~~~
V_{a}{}^{b}(z)\bt_{b}=\bt_{a}\x_{+}^{-1}\x_{-}\,,&\label{VxxV}\\
{}\nonumber\\
&R_{\mu}{}^{\nu}(z;g)\gamma_{\nu}
=\Omega(z;g)L^{-1}(z;g)\gamma_{\mu}L(z;g)\,,&\\
{}\nonumber\\
&\gamma^{\nu}R_{\nu}{}^{\mu}(z;g)
=\Omega(z;g)L(z;g)\gamma^{\mu}L^{-1}(z;g)\,.&\label{gR}
\end{subeqnarray}
where $R_{\mu}{}^{\nu}(z;g)$ is identical to the 
definition~(\ref{ehomogeneous}).   
We may  normalize $R_{\mu}{}^{\nu}(z;g)$  as
\begin{equation}
\hat{R}_{\mu}{}^{\nu}(z;g)=\Omega(z;g)^{-1}R_{\mu}{}^{\nu}(z;g)=\textstyle{\frac{1}{2}}\mbox{tr}(\gamma_{\mu}L(z;g)\gamma^{\nu}L^{-1}(z;g))\in \mbox{SO}(1,2)\,.
\label{normalize}
\end{equation}

\subsection{Representations\label{subrep}}
Based on the results in the previous subsection, it is easy to show that  
the matrix, ${\cal R}_{M}{}^{N}(z;g)$, given in 
eq.(\ref{calR})  is of the form
\begin{equation}
{\cal R}_{M}{}^{N}(z;g)
=\left(
\begin{array}{cc}
\Omega(z;g)\hat{R}_{\mu}{}^{\nu}(z;g)&
i\Omega(z;g)^{\frac{1}{2}}(L^{-1}(z;g)\gamma_{\mu}\Sigma^{b}(z;g))^{\beta}\\
0&\Omega(z;g)^{\frac{1}{2}}L^{-1}{}^{\beta}{}_{\alpha}(z;g)U_{a}{}^{b}(z;g)
\end{array}\right)\,.
\label{calRR}
\end{equation}
Since  ${\cal R}_{M}{}^{N}(z;g)$ is a representation of the 
three-dimensional superconformal group,  
each of the following also forms a representation of the group, though it is  
not a faithful representation
\begin{equation}
\begin{array}{cc}
\Omega(z;g)\in\mbox{D}\,,~~~~&~~~~
\hat{R}(z;g)\in\mbox{SO}(1,2)\,,\\
{}&{}\\
L(z;g)\,,~~~~&~~~~U(z;g)\in\mbox{O}(\N)\,,
\end{array}
\end{equation}
where $\mbox{D}$ is the one dimensional group of dilations. \newline
Under the successive superconformal transformations,~$g^{\prime\prime}: 
z\stackrel{g}{\longrightarrow}
z^{\prime}\stackrel{g^{\prime}}{\longrightarrow}z^{\prime\prime}$, they satisfy
\begin{equation}
\begin{array}{cc}
L(z;g)L(z^{\prime};g^{\prime})=L(z;g^{\prime\prime})\,,~~~~&~~~~
\mbox{and~so~on.}
\end{array}
\end{equation}

\subsection{Functions of Two Points\label{F2}}
In this subsection,  we construct matrix  valued   
functions depending on  two points, $z_{1}$ and $z_{2}$, 
in superspace which transform covariantly like a product of two tensors 
at $z_{1}$ and $z_{2}$  under superconformal transformations.  \newline

\indent   If  $F(z)$ is defined for $z\in\Real^{3|2{\cal N}}$ by
\begin{equation}
F(z)=\bar{Z}(0)G_{T}(z)Z(0)
=\left(\begin{array}{rc}
              i\x_{-}&\sqrt{2}\theta^{b}\\
           -\sqrt{2}\bt_{a}& \delta_{a}{}^{b}\end{array}\right)\,,
\label{Fform}
\end{equation}
then  $F(z)$ satisfies
\begin{equation}
\begin{array}{ll}
F(-z)&=\left(\begin{array}{cc}
\gamma^{0}&0\\
0&1
\end{array}\right)F(z)^{\dagger}\left(\begin{array}{cc}
\gamma^{0}&0\\
0&1
\end{array}\right)
=\left(\begin{array}{cc}
\e^{-1}&0\\
0&1
\end{array}\right)F(z)^{t}\left(\begin{array}{cc}
\e&0\\
0&1
\end{array}\right)\\
{}&{}\\
{}&=\left(\begin{array}{rc}
              -i\x_{+}&-\sqrt{2}\theta^{b}\\
           \sqrt{2}\bt_{a}& \delta_{a}{}^{b}\end{array}\right)\,,
\label{F-form}
\end{array}
\end{equation}
and the superdeterminant of $F(z)$ is given by
\begin{equation}
\mbox{sdet}\:F(z)=-\det\x_{+}=x^{2}
+\textstyle{\frac{1}{4}}(\bt_{a}\theta^{a})^{2}\,.
\label{sdetZZ}
\end{equation}

\indent   We also note 
\begin{equation}
\left(\begin{array}{cc}
               1&0\\
                -i\sqrt{2}\bt_{a}\x^{-1}_{-}&1
\end{array}\right)F(z)
\left(\begin{array}{cc}
               1&i\sqrt{2}\x^{-1}_{-}\theta^{b}\\
               0&\,1
\end{array}\right)=
\left(\begin{array}{cc}
               i\x_{-}&0\\
               0&V_{a}{}^{b}(-z)
\end{array}\right)\,,
\label{XV}
\end{equation}
where $V_{a}{}^{b}(-z)$ is identical to eq.(\ref{V-z}) and from eqs.(\ref{sdetZZ},\,\ref{XV}) it is evident that
\begin{equation}
\det V(z)=1\,.
\label{Vdet}
\end{equation}
Hence, with eq.(\ref{V-z}),  $V(z)\in\mbox{SO}(\N)$.\newline

\indent   Now with the  supersymmetric interval  for $\Real^{3|2{\cal N}}$ 
defined by 
\begin{equation}
\begin{array}{cc}
G_{T}(z_{2})^{-1}G_{T}(z_{1})=G_{T}(z_{12})\,,~~~~&~~~~
z^{M}_{12}=(x^{\mu}_{12},\theta^{a}_{12},\bt_{12a})=
-z^{M}_{21}\,,\\
{}&{}\\
x_{12}^{\mu}=x_{1}^{\mu}-x_{2}^{\mu}-i\bt_{2a}\gamma^{\mu}\theta_{1}^{a}\,,
~~~~&~~~~\theta^{a}_{12}=\theta^{a}_{1}-\theta^{a}_{2}\,,
\label{susyint}
\end{array}
\end{equation}
we may write
\begin{equation}
\displaystyle{\bar{Z}(z_{2})Z(z_{1})=F(z_{12})=
              \left(\begin{array}{rc}
              i\x_{12-}&\sqrt{2}\theta^{b}_{12}\\
                -\sqrt{2}\bt_{12a}& \delta_{a}{}^{b}\end{array}\right)}\,,
\label{ZZ}
\end{equation}
and 
\begin{equation}
\begin{array}{cc}
\mbox{sdet}\:F(z_{12})=x^{2}_{12}
+\textstyle{\frac{1}{4}}(\bt_{12a}\theta_{12}^{a})^{2}\,,~~~~&~~~~
\det V(z_{12})=1\,,
\end{array}
\end{equation}
where 
\begin{equation}
\begin{array}{c}
\x_{12-}=\x_{1-}-\x_{2+}-2i\theta^{a}_{2}\bt_{1a}
=\x_{12}-i\textstyle{\frac{1}{2}}\bt_{12a}\theta^{a}_{12}\,1\,,\\
{}\\
\x_{12+}=\x_{1+}-\x_{2-}+2i\theta^{a}_{1}\bt_{2a}
=\x_{12}+i\textstyle{\frac{1}{2}}\bt_{12a}\theta^{a}_{12}\,1\,.
\end{array}
\end{equation}

\indent   From eqs.(\ref{Ztr},\,\ref{barZtr})  $F(z_{12})$ transforms as
\begin{equation}
F(z^{\prime}_{12})
=\bar{K}(z_{2};g)F(z_{12})K(z_{1};g)\,.
\label{zz}
\end{equation}
Explicitly  with eqs.(\ref{Kform},\,\ref{Kbarform}) we get the 
transformation rules for $\x^{\prime}_{12\pm}$ and 
$\theta^{\prime a}_{12}$
\begin{subeqnarray}
\label{1212}
&\x^{\prime}_{12-}=\Omega(z_{1};g)^{\frac{1}{2}}\Omega(z_{2};g)^{\frac{1}{2}}
L^{-1}(z_{2};g)\x_{12-}L(z_{1};g)\,,&\nonumber\\
{}\label{green2}\\
&\x^{\prime}_{12+}=\Omega(z_{1};g)^{\frac{1}{2}}\Omega(z_{2};g)^{\frac{1}{2}}
L^{-1}(z_{1};g)\x_{12+}L(z_{2};g)\,,&\nonumber\\
{}\nonumber\\
&\theta^{\prime a}_{12}=\Omega(z_{1};g)^{\frac{1}{2}}L^{-1}(z_{1};g)(\theta_{12}^{b}U_{b}{}^{a}(z_{2};g)+i\x_{12+}\Sigma_{2}^{a})\,,&\nonumber\\
{}\label{theta12}\\
{}&\theta^{\prime a}_{21}=\Omega(z_{2};g)^{\frac{1}{2}}L^{-1}(z_{2};g)
(\theta_{21}^{b}U_{b}{}^{a}(z_{1};g)-i\x_{12-}\Sigma_{1}^{a})\,.&\nonumber
\end{subeqnarray}
In particular
\begin{equation}
x_{12}^{\prime 2}+\textstyle{\frac{1}{4}}(\bt_{12a}^{\prime}\theta_{12}^{\prime a})^{2}
=\Omega(z_{1};g)\Omega(z_{2};g)
(x_{12}^{2}+\textstyle{\frac{1}{4}}(\bt_{12a}\theta_{12}^{a})^{2})\,.
\label{denominator}
\end{equation}
From eqs.(\ref{gR},\,\ref{green2}) 
$\mbox{tr}(\gamma^{\mu}\x_{12-}\gamma^{\nu}\x_{12+})$ 
transforms covariantly as
\begin{equation}
\mbox{tr}(\gamma^{\mu}\x^{\prime}_{12-}\gamma^{\nu}\x^{\prime}_{12+})
=\mbox{tr}(\gamma^{\lambda}\x_{12-}\gamma^{\rho}\x_{12+})
R_{\lambda}{}^{\mu}(z_{2};g)R_{\rho}{}^{\nu}(z_{1};g)\,.
\label{green5}
\end{equation}
From eq.(\ref{theta12}) we get
\begin{subeqnarray}
\begin{array}{c}
{\left({\begin{array}{cc}
               1& -i\sqrt{2}\x^{-1}_{12-}\theta_{12}^{c}\\
               0&\delta_{a}{}^{c}
\end{array}}\right)K(z_{1};g)
\left({\begin{array}{cc}
               1& i\sqrt{2}\x^{\prime -1}_{12-}\theta_{12}^{\prime b}\\
               0&\delta_{d}{}^{b}
\end{array}}\right)}\\
{}\\
=\left({\begin{array}{cc}
               \Omega(z_{1};g)^{\frac{1}{2}}L(z_{1};g)&0\\
                0&U(z_{1};g)
\end{array}}\right)\,,~~\end{array}\\
{}\nonumber\\
{}\nonumber\\
\begin{array}{c}
{\left({\begin{array}{cc}
               1&0\\
                -i\sqrt{2}\bt^{\prime}_{12a}\x^{\prime -1}_{12-}&
\delta_{a}{}^{c}
\end{array}}\right)\bar{K}(z_{2};g)
\left({\begin{array}{cc}
               1&0\\
              i\sqrt{2}\bt_{12d}\x^{-1}_{12-}&\delta_{d}{}^{b}
\end{array}}\right)}\\
{}\\
=\left({\begin{array}{cc}
               \Omega(z_{2};g)^{\frac{1}{2}}L^{-1}(z_{2};g)&0\\
                0&U^{-1}(z_{2};g)
\end{array}}\right)\,.\end{array}
\end{subeqnarray}
Using this and eq.(\ref{XV}) we can rederive eq.(\ref{green2}) and obtain
\begin{equation}
\begin{array}{c}
V(z_{12}^{\prime})=U^{-1}(z_{1};g)V(z_{12})U(z_{2};g)\,,\\
{}\\
V(z_{21}^{\prime})=U^{-1}(z_{2};g)V(z_{21})U(z_{1};g)\,.
\end{array}
\label{green3}
\end{equation}

\subsection{Functions of Three Points}
In this subsection,  for three points, $z_{1},z_{2},z_{3}$ in superspace,   
we construct     `tangent' vectors, 
$\Z_{i}$, which transform homogeneously at $z_{i},\,i=1,2,3$. \newline
\indent  With $z_{21}\stackrel{i_{s}}{\longrightarrow}(z_{21})^{\prime},\,
z_{31}\stackrel{i_{s}}{\longrightarrow}(z_{31})^{\prime}$,   
we define $\Z_{1}^{M}=(X^{\mu}_{1},\Theta^{a}_{1})\in\Real^{3|2{\cal N}}$ by
\begin{equation}
G_{T}((z_{31})^{\prime})^{-1}{}G_{T}((z_{21})^{\prime})=G_{T}(\Z_{1})\,.
\label{F3defnew}
\end{equation}
Explicit expressions for $\Z^{M}_{1}$ can be obtained by calculating
\begin{equation}
\bar{Z}((z_{31})^{\prime})Z((z_{21})^{\prime})=F(\Z_{1})=
\left(\begin{array}{rc}
i\X_{1-}&\sqrt{2}\Theta^{b}_{1}\\
-\sqrt{2}\bar{\Theta}_{1a}& \delta_{a}{}^{b}
\end{array}\right)\,.
\label{F3def}
\end{equation}
We get 
\begin{equation}
\begin{array}{cc}
\multicolumn{2}{c}{
\X_{1-}=\x^{-1}_{31+}\x_{23-}\x_{21-}^{-1}\,,}\\
{}&{}\\
\Theta^{a}_{1}=i(\x_{21+}^{-1}\theta_{21}^{a}-\x_{31+}^{-1}\theta_{31}^{a})\,,
~~~~&~~~~
\bar{\Theta}_{1a}=-i(\bt_{21a}\x^{-1}_{21-}-\bt_{31a}\x^{-1}_{31-})\,.
\end{array}
\label{W1}
\end{equation}
Using
\begin{equation}
\x_{23-}=\x_{21-}-\x_{31+}-2i\theta_{31}^{a}\bt_{21a}\,,
\end{equation}
one can assure  
\begin{equation}
\begin{array}{l}
\X_{1+}=\gamma^{0}\X_{1-}^{\dagger}\gamma^{0}=-\e^{-1}\X_{1-}^{t}\e=
\x^{-1}_{21+}\x_{23+}\x^{-1}_{31-}\,,\\
{}\\
\X_{1+}-\X_{1-}=-2i\Theta^{a}_{1}\bar{\Theta}_{1a}=
i\bar{\Theta}_{1a}\Theta^{a}_{1}\,1\,.
\end{array}
\label{Ypro}
\end{equation}
From eq.(\ref{1212}) under  superconformal transformations,   
$z\stackrel{g}{\longrightarrow}z^{\prime}$,   
$\X_{1\pm},\,\Theta^{a}_{1},\,\bar{\Theta}_{1a}$ transform as
\begin{subeqnarray}
\label{X3transf}
&\X^{\prime}_{1\pm}=\Omega(z_{1};g)^{-1}L^{-1}(z_{1};g)
\X_{1\pm}L(z_{1};g)\,,&\\
{}\nonumber\\
&\Theta^{\prime a}_{1}=\Omega(z_{1};g)^{-\frac{1}{2}}L^{-1}(z_{1};g)
\Theta^{b}_{1}U_{b}{}^{a}(z_{1};g)\,,&\\
{}\nonumber\\
&\bT_{1a}^{\prime}=\Omega(z_{1};g)^{-\frac{1}{2}}
U^{-1}{}_{a}{}^{b}(z_{1};g)\bT_{1b}L(z_{1};g)\,,&
\end{subeqnarray}
so that
\begin{equation}
X_{1}^{\prime\mu}
=\Omega(z_{1};g)^{-1}X_{1}^{\nu}\hat{R}_{\nu}{}^{\mu}(z_{1};g)\,.
\label{XXXR}
\end{equation}
Thus $\Z_{1}$ transforms homogeneously at $z_{1}$,  
as    `tangent' vectors do. \newline
Eq.(\ref{X3transf}) can be summarized as
\begin{equation}
F(\Z_{1}^{\prime})=\left(\begin{array}{cc}
\Omega(z_{1};g)^{-\frac{1}{2}}L^{-1}(z_{1}g)&0\\
0&U^{-1}(z_{1};g)\end{array}\right)F(\Z_{1})
\left(\begin{array}{cc}
\Omega(z_{1};g)^{-\frac{1}{2}}L(z_{1}g)&0\\
0&U(z_{1};g)\end{array}\right)\,.
\end{equation}
Direct calculation using eq.(\ref{VxxV}) shows that
\begin{equation}
V(\Z_{1})=V(z_{12})V(z_{23})V(z_{31})\,.
\label{Vdecom}
\end{equation} 
Similarly for $R_{\mu}{}^{\nu}(z;i_{s})$ given in eq.(\ref{Rzi}) we obtain 
from eqs.(\ref{Rgamma},\ref{W1})
\begin{equation}
R(\Z_{1};i_{s})=
\left(x^{2}_{12}
+\textstyle{\frac{1}{4}}(\bt_{12a}\theta^{a}_{12})^{2}\right)^{2}
\left(x^{2}_{31}
+\textstyle{\frac{1}{4}}(\bt_{31a}\theta^{a}_{31})^{2}\right)^{2}
\,R(z_{12};i_{s})R(z_{23};i_{s})R(z_{31};i_{s})\,.
\label{1311223}
\end{equation}
From  eqs.(\ref{green5},\,\ref{green3})  
$V_{a}{}^{b}(\Z_{1}),R_{\mu}{}^{\nu}(\Z_{1};i_{s})$ 
transform homogeneously at $z_{1}$  
under superconformal transformation, 
$z\stackrel{g}{\longrightarrow}z^{\prime}$, 
\begin{subeqnarray}
&V(\Z_{1}^{\prime})=U^{-1}(z_{1};g)V(\Z_{1})U(z_{1};g)\,,&\\
&{}&\nonumber\\
&R(\Z^{\prime}_{1};i_{s})=\Omega(z_{1};g)^{2}
R^{-1}(z_{1};g)R(\Z_{1};i_{s})R(z_{1};g)\,.&
\end{subeqnarray}

\indent It is useful to note
\begin{equation}
\det \X_{1\pm}=-X_{1}^{2}-\textstyle{\frac{1}{4}}
(\bar{\Theta}_{1a}\Theta_{1}^{a})^{2}=-\displaystyle{\frac{
x_{23}^{2}+\textstyle{\frac{1}{4}}(\bar{\theta}_{23a}\theta_{23}^{a})^{2}}{
\left(x_{12}^{2}+\textstyle{\frac{1}{4}}(\bar{\theta}_{12a}\theta_{12}^{a})^{2}\right)\left(x_{31}^{2}+\textstyle{\frac{1}{4}}(\bar{\theta}_{31a}\theta_{31}^{a})^{2}\right)}}\,.
\label{detX}
\end{equation}

\indent  By taking cyclic permutations of   
$z_{1},z_{2},z_{3}$ in eq.(\ref{W1})  
we may define $\Z_{2},\Z_{3}$.    
We find $\Z_{2},\Z_{3}$ are related to $\Z_{1}$ as
\begin{subeqnarray}
\label{Z2Z3}
&\widetilde{\X_{2}}_{-}=-\x_{21+}\X_{1+}\x_{12+}\,,~~~~~~~~~
\widetilde{\Theta_{2}}^{a}=i\x_{21+}\Theta_{1}^{b}V_{b}{}^{a}(z_{12})\,,&\\
{}\nonumber\\
&\X_{3-}=\x_{31-}^{-1}\widetilde{\X_{1}}_{+}\x_{13-}^{-1}\,,~~~~~~~~~\,
\Theta^{a}_{3}=i\x_{31-}^{-1}\widetilde{\Theta_{1}}^{b}V_{b}{}^{a}(z_{13})\,,&
\end{subeqnarray}
where $\widetilde{\Z}=(\widetilde{X},\widetilde{\Theta})$ is defined by 
superinversion, $\Z\stackrel{i_{s}}{\longrightarrow}\widetilde{\Z}$.

\newpage

\section{Superconformal Invariance of Correlation Functions
\label{correlation}}
In this section  we discuss the    
superconformal invariance of correlation 
functions for  quasi-primary superfields and exhibit general forms of   
two-point, three-point and $n$-point functions without proof, as 
the proof is essentially identical to those 
in our earlier work~\cite{paper3,paper1}.

\subsection{Quasi-primary Superfields}
We first   assume  that there exist quasi-primary 
 superfields, 
$\Psi^{I}(z)$,  which   under the superconformal transformation, 
$z\stackrel{g}{\longrightarrow}z^{\prime}$,  transform as
\begin{equation}
\begin{array}{cc}
\Psi^{I}\longrightarrow\Psi^{\prime}{}^{I}\,,~~~&~~~ 
\Psi^{\prime}{}^{I}(z^{\prime})
=\Psi^{J}(z)D^{~I}_{J}(z;g)\,.
\end{array}
\label{primary}
\end{equation}
$D(z;g)$ obeys the group property   
so that under the 
successive superconformal transformations,~$g^{\prime\prime}: 
z\stackrel{g}{\longrightarrow}
z^{\prime}\stackrel{g^{\prime}}{\longrightarrow}z^{\prime\prime}$, it 
satisfies 
\begin{equation}
D(z;g)D(z^{\prime};g^{\prime})=D(z;g^{\prime\prime})\,,
\label{Drep}
\end{equation}
and hence 
\begin{equation}
D(z;g)^{-1}=D(z^{\prime};g^{-1})\,.
\end{equation}
We choose here $D(z;g)$ to be a representation of 
$\mbox{SO}(1,2)\times\mbox{O}({\cal N})\times\mbox{D}$, 
which is a subgroup of the stability group at $z=0$,   
and so  we decompose the spin index, $I$, of superfields into 
$\mbox{SO}(1,2)$ index, $\rho$, and $\mbox{O}({\cal N})$ index, $r$, as 
$\Psi^{I}\equiv \Psi_{\rho}{}^{r}$. Now  $D_{J}{}^{I}(z;g)$  
is factorized as 
\begin{equation}
D_{J}{}^{I}(z;g)=D^{\rho}{}_{\sigma}(L(z;g))D_{r}{}^{s}(U(z;g))
\Omega(z;g)^{-\eta}\,,
\label{DLU}
\end{equation}
where 
$D^{\rho}{}_{\sigma}(L),\,D_{r}{}^{s}(U)$ are  representations of 
$\mbox{SO}(1,2)$ and $\mbox{O}({\cal N})$ respectively,  while    
$\eta$ is the scale dimension  of $\Psi_{\rho}{}^{r}$. \newline

\indent Infinitesimally 
\begin{equation}
\delta\Psi_{\rho}{}^{r}(z)=
-({\cal L}+\eta\hat{\lambda}(z))\Psi_{\rho}{}^{r}(z)
-\Psi_{\sigma}{}^{r}(z)(s^{\alpha}{}_{\beta})^{\sigma}{}_{\rho}\hat{w}^{\beta}{}_{\alpha}(z)
-\Psi_{\rho}{}^{s}(z)\textstyle{\frac{1}{2}}(s_{ab})_{s}{}^{r}
\hat{t}^{ab}(z)\,,
\label{deltaPsi}
\end{equation}
where $\hat{t}^{ab}(z)=\delta^{ac}\hat{t}_{c}{}^{b}(z)$, and 
$s^{\alpha}{}_{\beta},\,s_{ab}$ satisfy
\begin{equation}
\begin{array}{c}
[s^{\alpha}{}_{\beta},s^{\gamma}{}_{\delta}]
=\delta^{\alpha}{}_{\delta}s^{\gamma}{}_{\beta}-\delta_{\beta}{}^{\gamma}s^{\alpha}{}_{\delta}\,,\\
{}\\
{}[s_{ab},s_{cd}]=-\eta_{ac}s_{bd}+\eta_{ad}s_{bc}+\eta_{bc}s_{ad}
-\eta_{bd}s_{ac}\,.
\end{array}
\label{stcom1}
\end{equation}
$s_{ab}$ is the  generator  of $\mbox{O}({\cal N})$, while  
$s^{\alpha}{}_{\beta}$ is connected to the generator of $\mbox{SO}(1,2)$,  
$s_{\mu\nu}$,   through
\begin{equation}
\begin{array}{cc}
s_{\mu\nu}\equiv\textstyle{\frac{1}{2}}s^{\alpha}{}_{\beta}
(\gamma_{[\mu}\gamma_{\nu]})^{\beta}{}_{\alpha}\,,~~~~&~~~~
s^{\alpha}{}_{\beta}=-\textstyle{\frac{1}{2}}s_{\mu\nu}
(\gamma^{[\mu}\gamma^{\nu]})^{\alpha}{}_{\beta}\,,\\
{}&{}\\
\multicolumn{2}{c}{
[s_{\mu\nu},s_{\lambda\rho}]=-\eta_{\mu\lambda}s_{\nu\rho}+\eta_{\mu\rho}
s_{\nu\lambda}+\eta_{\nu\lambda}s_{\mu\rho}-\eta_{\nu\rho}s_{\mu\lambda}\,,}\\
{}&{}\\
\multicolumn{2}{c}{
s^{\alpha}{}_{\beta}\hat{w}^{\beta}{}_{\alpha}(z)=
\textstyle{\frac{1}{2}}s_{\mu\nu}\hat{w}^{\mu\nu}(z)\,.}
\end{array}
\label{stcom2}
\end{equation}
\newline
\indent From eqs.(\ref{defL3},\,\ref{Kcom}) using eq.(\ref{stcom1}) we have
\begin{equation}
\delta_{3}\Psi_{\rho}{}^{r}=[\delta_{2},\delta_{1}]\Psi_{\rho}{}^{r}\,.
\end{equation}
\indent  
It is useful to consider the conjugate superfield of $\Psi_{\rho}{}^{r}$, 
$\bar{\Psi}^{\rho}{}_{r}(z)$, which  transforms as
\begin{equation}
\bar{\Psi}^{\prime\rho}{}_{r}(z^{\prime})=\Omega(z;g)^{-\eta}
D^{\rho}{}_{\sigma}(L^{-1}(z;g))D_{r}{}^{s}(U^{-1}(z;g))
\bar{\Psi}^{\sigma}{}_{s}(z)\,.
\end{equation} 
\indent  Superconformal invariance for a general $n$-point  function requires
\begin{equation}
\langle
\Psi_{1}^{\prime I_{1}}(z_{1})\Psi_{2}^{\prime I_{2}}(z_{2}) 
\cdots\Psi_{n}^{\prime I_{n}}(z_{n})\rangle
=\langle \Psi_{1}^{I_{1}}(z_{1})\Psi_{2}^{I_{2}} (z_{2})
\cdots \Psi_{n}^{I_{n}}(z_{n})\rangle\,.
\label{Green}
\end{equation}

\subsection{Two-point Correlation Functions}
The  solution for 
the two-point  function of the quasi-primary superfields, 
$\Psi_{\rho}{}^{r},\bar{\Psi}^{\rho}{}_{r}$,  has the general  form
\begin{equation}
\displaystyle{
\langle \bar{\Psi}^{\rho}{}_{r}(z_{1})\Psi_{\sigma}{}^{s}(z_{2})\rangle = 
C_{\Psi}\frac{I^{\rho}{}_{\sigma}(\hat{\x}_{12+})I_{r}{}^{s}(V(z_{12}))}{
\left(x^{2}_{12}+\textstyle{\frac{1}{4}}(\bt_{12a}\theta^{a}_{12})^{2}\right)^{\eta}}\,,}
\label{2gen}
\end{equation}
where we put 
\begin{equation}
\displaystyle{\hat{\x}_{12+}=
\frac{\x_{12+}}{\left(x^{2}_{12}+\textstyle{\frac{1}{4}}(\bt_{12a}\theta^{a}_{12})^{2}\right)^{\frac{1}{2}}}\,,}
\end{equation}
and   $I^{\rho}{}_{\sigma}(\hat{\x}_{12+}),\,I_{r}{}^{s}(V(z_{12}))$ 
are   tensors transforming covariantly according to the    
appropriate representations of $\mbox{SO}(1,2),\,\mbox{O}({\cal N})$ 
which  are formed by decomposition of tensor products of   
$\hat{\x}_{12+},\,V(z_{12})$. Under superconformal transformations,  
$I^{\rho}{}_{\sigma}(\hat{\x}_{12+})$ and  $I_{r}{}^{s}(V(z_{12}))$ satisfy from 
eqs.(\ref{green2},\,\ref{green3})
\begin{subeqnarray}
\label{IIpro}
&D(L^{-1}(z_{1};g))I(\hat{\x}_{12+})D(L(z_{2};g))
=I(\hat{\x}{}^{\prime}_{12+})\,,&\\
{}\nonumber\\
&D(U^{-1}(z_{1};g))I(V(z_{12}))D(U(z_{2};g))=I(V(z^{\prime}_{12}))\,.&
\end{subeqnarray}
\newline
\indent  As  examples,  we first consider  real scalar, spinorial and gauge  
superfields,    \\
 $S(z),\,\phi_{\alpha}(z),\,\bar{\phi}^{\alpha}(z),\,\zeta^{a}(z),\,
\zeta_{a}(z)$. They satisfy
\begin{equation}
\begin{array}{l}
S(z)~=S(z)^{\ast}\,,\\
{}\\
\bar{\phi}^{\alpha}(z)=\e^{-1\alpha\beta}\phi_{\beta}(z)=(\gamma^{0}\phi(z)^{\dagger})^{\alpha}\,,\\
{}\\
\zeta^{a}(z)=\zeta^{a}(z)^{\ast}=\zeta_{a}(z)\,,
\end{array}
\end{equation}
and transform as
\begin{equation}
\begin{array}{l}
S^{\prime}(z^{\prime})~=\Omega(z;g)^{-\eta}S(z)\,,\\
{}\\
\phi^{\prime}_{\alpha}(z^{\prime})=\Omega(z;g)^{-\eta}
\phi_{\beta}(z)L^{\beta}{}_{\alpha}(z;g)\,,\\
{}\\
\bar{\phi}^{\prime\alpha}(z^{\prime})=\Omega(z;g)^{-\eta}
L^{-1}{}^{\alpha}{}_{\beta}(z;g)\bar{\phi}^{\beta}(z)\,,\\
{}\\
\zeta^{\prime a}(z^{\prime})=\Omega(z;g)^{-\eta}\zeta^{b}(z)
U_{b}{}^{a}(z;g)\,,\\
{}\\
\zeta^{\prime}_{a}(z^{\prime})=\Omega(z;g)^{-\eta}U^{-1}{}_{a}{}^{b}(z;g)
\zeta_{b}(z)
\,.
\end{array}
\end{equation}
The two-point functions of  them are 
\begin{eqnarray}
&\displaystyle{
\langle S(z_{1})S(z_{2})\rangle = 
C_{S}\frac{1}{\left(x^{2}_{12}+\textstyle{\frac{1}{4}}(\bt_{12a}\theta^{a}_{12})^{2}\right)^{\eta}}\,,}&\\
{}\nonumber\\
&\displaystyle{
\langle \bar{\phi}^{\alpha}(z_{1})\phi_{\beta}(z_{2})\rangle =
iC_{\phi}\frac{(\x_{12+})^{\alpha}{}_{\beta}}{
\left(x^{2}_{12}+\textstyle{\frac{1}{4}}(\bt_{12a}\theta^{a}_{12})^{2}\right)^{\eta+\frac{1}{2}}}\,,}&\\
{}\nonumber\\
&\displaystyle{
\langle \zeta_{a}(z_{1})\zeta^{b}(z_{2})\rangle =
C_{\zeta}\frac{V_{a}{}^{b}(z_{12})}{
\left(x^{2}_{12}+\textstyle{\frac{1}{4}}(\bt_{12a}\theta^{a}_{12})^{2}
\right)^{\eta}}\,.}&\\
\end{eqnarray}
Note that  to have
non-vanishing two-point correlation functions,  the 
 scale dimensions,$\,\eta$, 
of the two fields must be equal. \newline

\indent  For a  real vector superfield, $J^{\mu}(z)$,  where the representation
of $\mbox{SO}(1,2)$  is given by
$\hat{R}_{\mu}{}^{\nu}(z;g)$,  we have
\begin{equation}
\displaystyle{
\langle J^{\mu}(z_{1})J^{\nu}(z_{2})\rangle =
C_{V}\frac{I^{\mu\nu}(z_{12})}{\left(x^{2}_{12}+\textstyle{\frac{1}{4}}(\bt_{12a}\theta^{a}_{12})^{2}
\right)^{\eta}}}\,,
\label{VV}
\end{equation}
where
\begin{equation}
I^{\mu\nu}(z)=\hat{R}^{\mu\nu}(z;i_{s})
=\textstyle{\frac{1}{2}}\mbox{tr}(
\gamma^{\mu}\hat{\x}_{+}\gamma^{\nu}\hat{\x}_{-})\,.
\end{equation}
From eq.(\ref{Rgamma}) we note
\begin{equation}
\begin{array}{cc}
I^{\mu\nu}(z)=I^{\nu\mu}(-z)\,,~~~~&~~~~
I^{\mu\nu}(z)I_{\lambda\nu}(z)=\delta^{\mu}{}_{\lambda}\,.
\end{array}
\end{equation}
If we define
\begin{equation}
J^{\alpha}{}_{\beta}(z)=J^{\mu}(z)(\gamma_{\mu})^{\alpha}{}_{\beta}\,,
\end{equation}
then from eqs.(\ref{complete},\ref{VV}) and
\begin{equation}
\begin{array}{l}
D_{a\omega}(z_{1})(\x_{12+})^{\alpha}{}_{\beta}=2i
\delta_{\omega}{}^{\alpha}\bt_{12 a\beta}\,,\\
{}\\
D_{a\omega}(z_{1})(\x_{12-})^{\alpha}{}_{\beta}=2i
(\delta_{\omega}{}^{\alpha}\bt_{12 a\beta}-
\delta^{\alpha}{}_{\beta}\bt_{12 a\omega})\,,\\
{}\\
D_{a\omega}(z_{1})(x^{2}_{12}
+\textstyle{\frac{1}{4}}(\bt_{12a}\theta^{a}_{12})^{2})=
2i(\bt_{12a}\x_{12-})_{\alpha}\,,
\end{array}
\end{equation}
we get
\begin{equation}
\displaystyle{
D_{a\alpha}(z_{1})
\langle J^{\alpha}{}_{\beta}(z_{1})J^{\nu}(z_{2})\rangle =2i
C_{V}(2-\eta)\frac{(\bt_{12a}\gamma^{\nu}\x_{12-})_{\beta}}{
\left(x^{2}_{12}+\textstyle{\frac{1}{4}}(\bt_{12a}\theta^{a}_{12})^{2}
\right)^{\eta +1}}}\,.
\end{equation}
Hence $\langle J^{\alpha}{}_{\beta}(z_{1})J^{\nu}(z_{2})\rangle$ is conserved if 
$\eta=2$
\begin{equation}
\begin{array}{cc}
D_{a\alpha}(z_{1})\langle J^{\alpha}{}_{\beta}(z_{1})J^{\nu}(z_{2})\rangle=0~~~~~~&~~~~~~\mbox{if~~}\eta=2\,.
\end{array}
\label{cons1}
\end{equation}
The anti-commutator relation for  
$D_{a\alpha}$~(\ref{anticomD}) implies also 
\begin{equation}
\begin{array}{cc}
\displaystyle{
\frac{\partial~}{\partial x^{\mu}_{1}}}
\langle J^{\mu}(z_{1})J^{\nu}(z_{2})\rangle=0~~~~~~&~~~~~~~~~\mbox{if~~}\eta=2\,.
\end{array}
\label{cons2}
\end{equation}
This agrees with the non-supersymmetric general result that 
two-point correlation function of vector field in $d$-dimensional 
conformal theory is conserved if the scale dimension is $d-1$~\cite{hughpetkou}.

\subsection{Three-point Correlation Functions}
The  solution for the three-point correlation  function of the quasi-primary
superfields,~$\Psi_{\rho}{}^{r}$, 
has the general  form
\begin{equation}
\begin{array}{l}
\displaystyle{
\langle \Psi_{1\rho}{}^{r}(z_{1})\Psi_{2\sigma}{}^{s}(z_{2})
\Psi_{3\tau}{}^{t}(z_{3})\rangle }\\
{}\\
=\displaystyle{ 
\frac{H_{\rho}{}^{r}{}_{\sigma^{\prime}}{}^{s^{\prime}}{}_{\tau^{\prime}}{}^{
t^{\prime}}(\Z_{1})I^{\sigma^{\prime}}{}_{\sigma}(\hat{\x}_{12+})
I^{\tau^{\prime}}{}_{\tau}(\hat{\x}_{13+})I_{s^{\prime}}{}^{s}(V(z_{12}))
I_{t^{\prime}}{}^{t}(V(z_{13}))}
{\left(x^{2}_{12}+\textstyle{\frac{1}{4}}(\bt_{12a}\theta^{a}_{12})^{2}\right)^{\eta_{2}}\left(x^{2}_{13}+\textstyle{\frac{1}{4}}(\bt_{13a}\theta^{a}_{13})^{2}\right)^{\eta_{3}}}}\,,
\end{array}
\label{3gen}
\end{equation}
where  $\Z_{1}{}^{M}
=(X_{1}^{\mu},\Theta_{1}^{a})\in\Real^{3|2{\cal N}}$ is given by 
eq.(\ref{F3defnew}). \newline

\indent Superconformal invariance~(\ref{Green}) is  now equivalent to   
\begin{subeqnarray}
\label{equiv}
\begin{array}{c}
H_{\rho^{\prime}}{}^{r}{}_{\sigma^{\prime}}{}^{s}{}_{\tau^{\prime}}
{}^{t}(\Z)D^{\rho^{\prime}}{}_{\rho}(L)D^{\sigma^{\prime}}{}_{\sigma}(L)
D^{\tau^{\prime}}{}_{\tau}(L)=H_{\rho}{}^{r}{}_{\sigma}{}^{s}{}_{\tau}
{}^{t}(\Z^{\prime})\,,\\
{}\\
\Z^{\prime M}=(X^{\nu}\hat{R}_{\nu}{}^{\mu}(L),\,L^{-1}\Theta^{a})\,,
\end{array}\\
{~}\nonumber\\
{~}\nonumber\\
\begin{array}{c}
H_{\rho}{}^{r^{\prime}}{}_{\sigma}{}^{s^{\prime}}{}_{\tau}{}^{t^{\prime}}(\Z)
D_{r^{\prime}}{}^{r}(U)D_{s^{\prime}}{}^{s}(U)D_{t^{\prime}}{}^{t}(U)
=H_{\rho}{}^{r}{}_{\sigma}{}^{s}{}_{\tau}
{}^{t}(\Z^{\prime\prime})\,,\\
{}\\
\Z^{\prime\prime M}=(X^{\mu},\,\Theta^{b}U_{b}{}^{a})\,,
\end{array}\\
{~}\nonumber\\
{~}\nonumber\\
\begin{array}{c}
H_{\rho}{}^{r}{}_{\sigma}{}^{s}{}_{\tau}{}^{t}(\Z)=
\lambda^{\eta_{2}+\eta_{3}-\eta_{1}}
H_{\rho}{}^{r}{}_{\sigma}{}^{s}{}_{\tau}{}^{t}(\Z^{\prime\prime\prime})\,,\\
{}\\
\Z^{\prime\prime\prime M}=(\lambda X^{\mu},\,\lambda^{\frac{1}{2}}\Theta^{a})\,,
\end{array}~~~~~~~~~~
\end{subeqnarray}
where $U\in\mbox{O}({\cal N}),\,\lambda\in\Real$ and $2\times 2$ matrix, $L$, 
satisfies
\begin{equation}
\begin{array}{cc}
L^{-1}=\gamma^{0}L^{\dagger}\gamma^{0}=\e^{-1}L^{t}\e\,,~~~~&~~~~\det L=1\,,\\
{}&{}\\
\multicolumn{2}{c}{\hat{R}_{\nu}{}^{\mu}(L)=\textstyle{\frac{1}{2}}
\mbox{tr}(\gamma_{\nu}L\gamma^{\mu}L^{-1})\,.}
\end{array}
\label{LRdef}
\end{equation}

\indent In general there are  a finite number of linearly independent  
solutions of eq.(\ref{equiv}), and this number can   be considerably     
reduced by taking into account the symmetry properties,  
superfield conservations and the superfield 
constraints~\cite{paper1,9808041,9907107}.  

\subsection{$n$-point Correlation Functions - in general\label{subn}}
The solution for  $n$-point  correlation functions  of the quasi-primary superfields,~$\Psi_{\rho}{}^{r}$,  has the general  form  
\begin{equation}
\begin{array}{l}
\langle \Psi_{1\rho_{1}}{}^{r_{1}}(z_{1})\cdots\Psi_{n\rho_{n}}{}^{r_{n}}(z_{n})\rangle\\
{}\nonumber\\
=H_{\rho_{1}}{}^{r_{1}}{}_{\rho_{2}^{\prime}}{}^{r_{2}^{\prime}}\cdots
{}_{\rho_{n}^{\prime}}{}^{r_{n}^{\prime}}(\Z_{1(1)},\cdots,\Z_{1(n-2)})
\displaystyle{\prod_{k=2}^{n}
\frac{I^{\rho^{\prime}_{k}}{}_{\rho_{k}}(\hat{\x}_{1k+})
I_{r^{\prime}_{k}}{}^{r_{k}}(V(z_{1k}))}{
\left(x^{2}_{1k}+\textstyle{\frac{1}{4}}(\bt_{1ka}\theta^{a}_{1k})^{2}\right)^{\eta_{k}}}\,,}
\end{array}
\label{ngen}
\end{equation}
where, in a similar fashion to eq.(\ref{F3defnew}),   with   
$z_{k1}\stackrel{i_{s}}{\longrightarrow}\widetilde{z_{k1}},\,k\geq 2$,    
$\Z_{1(1)},\cdots,\Z_{1(n-2)}$ are  given by
\begin{equation}
\begin{array}{cc}
G_{T}(\widetilde{z_{n1}})^{-1}{}G_{T}(\widetilde{z_{j1}})=G_{T}(\Z_{1(j-1)})\,,
~~~~~&~~~~
j=2,3,\cdots,n-1\,.
\end{array}
\end{equation}
We note that all of them are `tangent' vectors at $z_{1}$.\newline

\indent Superconformal invariance~(\ref{Green}) is  equivalent to
\begin{subeqnarray}
\label{equivgen}
&\begin{array}{c}
H_{\rho^{\prime}_{1}}{}^{r_{1}}\cdots
{}_{\rho_{n}^{\prime}}{}^{r_{n}}(\Z_{(1)},\cdots,\Z_{(n-2)})
\displaystyle{\prod_{k=1}^{n}}D^{\rho_{k}^{\prime}}{}_{\rho_{k}}(L)=
H_{\rho_{1}}{}^{r_{1}}\cdots
{}_{\rho_{n}}{}^{r_{n}}(\Z^{\prime}_{(1)},\cdots,\Z^{\prime}_{(n-2)})\,,\\
{}\\
\Z^{\prime M}_{(j)}=(X_{(j)}^{\nu}\hat{R}_{\nu}{}^{\mu}(L),\,
L^{-1}\Theta_{(j)}^{a})\,,
\end{array}&\label{equivgenL}\\
{}\nonumber\\
{}\nonumber\\
&\begin{array}{c}
H_{\rho_{1}}{}^{r^{\prime}_{1}}\cdots
{}_{\rho_{n}}{}^{r^{\prime}_{n}}(\Z_{(1)},\cdots,\Z_{(n-2)})
\displaystyle{\prod_{k=1}^{n}}D_{r_{k}^{\prime}}{}^{r_{k}}(U)=
H_{\rho_{1}}{}^{r_{1}}\cdots
{}_{\rho_{n}}{}^{r_{n}}(\Z^{\prime\prime}_{(1)},\cdots,\Z^{\prime\prime}_{(n-2)})\,,\\
{}\\
\Z_{(j)}^{\prime\prime M}=( X^{\mu}_{(j)},\,
\Theta_{(j)}^{b}U_{b}{}^{a})\,,
\end{array}&\label{equivgenO}\\
{}\nonumber\\
{}\nonumber\\
&\begin{array}{c}
H_{\rho_{1}}{}^{r_{1}}\cdots{}_{\rho_{n}}{}^{r_{n}}
(\Z_{(1)},\cdots,\Z_{(n-2)})=\lambda^{-\eta_{1}+\eta_{2}+\cdots+\eta_{n}}
H_{\rho_{1}}{}^{r_{1}}\cdots
{}_{\rho_{n}}{}^{r_{n}}(\Z^{\prime\prime\prime}_{(1)},\cdots,
\Z^{\prime\prime\prime}_{(n-2)})\,,\\
{}\\
\Z_{(j)}^{\prime\prime\prime M}=( \lambda
X^{\mu}_{(j)},\,
\lambda^{\frac{1}{2}}\Theta^{a}_{(j)})\,.
\end{array}&\label{equivgenD}
\end{subeqnarray}
Thus  $n$-point 
functions  reduce to one  unspecified $(n-2)$-point  function 
which must transform  homogeneously  under the 
rigid transformations,  
$\mbox{SO}(1,2)\times\mbox{O}({\cal N})\times\mbox{D}$.\newline

\indent  From
\begin{equation}
\begin{array}{cc}
\X_{1(j-1)+}=\x^{-1}_{j1+}\x_{jn+}\x_{n1-}^{-1}\,,~~~~&~~~~
\X_{1(j-1)-}=\x^{-1}_{n1+}\x_{jn-}\x_{j1-}^{-1}\,,
\end{array}
\end{equation}
we get
\begin{equation}
\X_{(l,m)+}\equiv\X_{1(l-1)+}-\X_{1(m-1)-}+2i\Theta_{1(l-1)}^{a}
\bar{\Theta}_{1(m-1)a}=\x_{l1+}^{-1}\x_{lm+}\x^{-1}_{m1-}\,,
\end{equation}
and hence
\begin{equation}
\displaystyle{\det \X_{(l,m)\pm}=-\frac{x_{lm}^{2}+\textstyle{\frac{1}{4}}
(\bt_{lma}\theta^{a}_{lm})^{2}}{\left(x_{1l}^{2}+\textstyle{\frac{1}{4}}
(\bt_{1la}\theta^{a}_{1l})^{2}\right)\left(x_{1m}^{2}+\textstyle{\frac{1}{4}}
(\bt_{1ma}\theta^{a}_{1m})^{2}\right)}}\,.
\label{forcross}
\end{equation}
Now if we define
\begin{equation}
\Delta_{lm}=-\textstyle{\frac{1}{2(n-1)(n-2)}}\displaystyle{
\sum_{i=1}^{n}\eta_{i}
+\textstyle{\frac{1}{2(n-2)}}(\eta_{l}+\eta_{m})}\,,
\end{equation}
then using the following identity which holds for any matrix, 
$S_{lm}$, and  number, $\lambda$, 
\begin{equation}
\displaystyle{\left(\prod_{l\neq m}
(S_{lm})^{\Delta_{lm}}\right)\left(\prod_{k=2}^{n}
(S_{1k}S_{k1})^{-\frac{1}{2}\eta_{k}}\right)=
\lambda^{-\frac{1}{2}(-\eta_{1}+\eta_{2}+\cdots+\eta_{n})}
\prod_{2\leq l\neq m}\left(\frac{\lambda S_{lm}}{S_{l1}S_{1m}}
\right)^{\Delta_{lm}}}\,,
\label{identity}
\end{equation}
we can rewrite the $n$-point  correlation functions~(\ref{ngen}) as 
\begin{equation}
\begin{array}{l}
\langle \Psi_{1\rho_{1}}{}^{r_{1}}(z_{1})\cdots\Psi_{n\rho_{n}}{}^{r_{n}}(z_{n})\rangle\\
{}\nonumber\\
=\displaystyle{\frac{
K_{\rho_{1}}{}^{r_{1}}{}_{\rho_{2}^{\prime}}{}^{r_{2}^{\prime}}\cdots
{}_{\rho_{n}^{\prime}}{}^{r_{n}^{\prime}}(\Z_{1(1)},\cdots,\Z_{1(n-2)})
\prod_{k=2}^{n}I^{\rho^{\prime}_{k}}{}_{\rho_{k}}(\hat{\x}_{1k+})
I_{r^{\prime}_{k}}{}^{r_{k}}(V(z_{1k}))}{
\prod_{l\neq m}
\left(x^{2}_{lm}+\textstyle{\frac{1}{4}}(\bt_{lma}\theta^{a}_{lm})^{2}\right)^{\Delta_{lm}}}}\,,
\end{array}
\label{ngen2}
\end{equation}
where
\begin{equation}
K_{\rho_{1}}{}^{r_{1}}\cdots{}_{\rho_{n}}{}^{r_{n}}(\Z_{1(1)},
\cdots,\Z_{1(n-2)})=
H_{\rho_{1}}{}^{r_{1}}\cdots{}_{\rho_{n}}{}^{r_{n}}(\Z_{1(1)},
\cdots,\Z_{1(n-2)})
\displaystyle{\prod_{2\leq l\neq m}}(-\det\X_{(l,m)\pm})^{\Delta_{lm}}\,.
\end{equation}
Note the difference in eq.(\ref{ngen}) and eq.(\ref{ngen2}),  namely  the 
denominator in the latter is written in a democratic fashion.\newline

\indent Superconformal invariance~(\ref{equivgen}) is  equivalent to
\begin{equation}
\begin{array}{c}
K_{\rho^{\prime}_{1}}{}^{r^{\prime}_{1}}\cdots
{}_{\rho_{n}^{\prime}}{}^{r^{\prime}_{n}}(\Z_{(1)},\cdots,\Z_{(n-2)})
\displaystyle{\prod_{k=1}^{n}}D^{\rho_{k}^{\prime}}{}_{\rho_{k}}(L)
D_{r^{\prime}_{k}}{}^{r_{k}}(U)=
K_{\rho_{1}}{}^{r_{1}}\cdots
{}_{\rho_{n}}{}^{r_{n}}(\Z^{\prime}_{(1)},\cdots,\Z^{\prime}_{(n-2)})\,,\\
{}\\
\Z^{\prime M}_{(j)}=(\lambda X_{(j)}^{\nu}\hat{R}_{\nu}{}^{\mu}(L),\,
\lambda^{\frac{1}{2}}L^{-1}\Theta_{(j)}^{b}U_{b}{}^{a})\,.
\end{array}
\label{equivgen2}
\end{equation}
In particular,  $K$ is invariant under dilations contrary to $H$.

\subsection{Superconformal Invariants\label{subn2}}
In the case of correlation functions of quasi-primary scalar superfields,  
eqs.(\ref{Green},\ref{ngen2},\ref{equivgen2}) imply that     
$K(\Z_{1(1)},\cdots,\Z_{1(n-2)})$ is a function of the superconformal 
invariants and furthermore that all of  the superconformal invariants     
can be generated  by contracting the indices of $\Z^{M}_{1(j)}=(X_{1(j)}^{\mu},
\Theta_{1(j)}^{a\alpha})$ to make them 
$\mbox{SO}(1,2)\times\mbox{O}({\cal N})\times\mbox{D}$      invariant according to the recipe by Weyl~\cite{weyl}. To do so we first normalize 
$\Z_{1(j)}^{\mu}$ as
\begin{equation}
\begin{array}{cc}
\multicolumn{2}{c}
{\hat{\Z}_{1(j)}^{\mu}=(\hat{X}_{1(j)}^{\mu},\hat{\Theta}_{1(j)}^{a})\,,}\\
{}&{}\\
\hat{X}^{\mu}_{1(j)}=\displaystyle{\frac{X^{\mu}_{1(j)}}{(X_{1(1)}^{2})^{\frac{1}{2}}}}\,,~~~~&~~~~
\hat{\Theta}^{a}_{1(j)}=\displaystyle{\frac{\Theta^{a}_{1(j)}}{(X_{1(1)}^{2})^{\frac{1}{4}}}}\,.
\end{array}
\end{equation}
By virtue of eqs.(\ref{completere},\ref{N=1sigma2})  all the   
$\mbox{SO}(1,2)\times\mbox{O}({\cal N})\times\mbox{D}$      invariants or three-dimensional superconformal invariants are
\begin{equation}
\begin{array}{cccc}
\hat{X}_{1(j)}{\cdot\hat{X}_{1(k)}}\,,~~~~&~~~\hat{\bar{\Theta}}_{1(j)a}\hat{X}_{1(k)}{\cdot\gamma}\hat{\Theta}_{1(l)}^{a}\,,~~~~&~~~
\hat{\bar{\Theta}}_{1(j)a}\hat{\Theta}_{1(k)}^{b}
\hat{\bar{\Theta}}_{1(l)b}\hat{\Theta}_{1(m)}^{a}\,,~~~~&~~~
\hat{\bar{\Theta}}_{1(j)a}\hat{\Theta}_{1(k)}^{a}\,.
\end{array}
\end{equation}
In particular, from eq.(\ref{forcross}) we note that  they produce 
cross ratio type invariants depending on four points,  
$z_{r},z_{s},z_{t},z_{u}$, 
the non-supersymmetric of  which are well known - see e.g. Ref. \cite{ginsparg}
\begin{equation}
\displaystyle{\frac{\left(x_{rs}^{2}+\textstyle{\frac{1}{4}}
(\bt_{rsa}\theta^{a}_{rs})^{2}\right)\left(x_{tu}^{2}+\textstyle{\frac{1}{4}}
(\bt_{tua}\theta^{a}_{tu})^{2}\right)}{\left(x_{ru}^{2}+\textstyle{\frac{1}{4}}
(\bt_{rua}\theta^{a}_{ru})^{2}\right)\left(x_{ts}^{2}+\textstyle{\frac{1}{4}}
(\bt_{tsa}\theta^{a}_{ts})^{2}\right)}}\,.
\end{equation}
\newline

\indent    
If we restrict the  $R$-symmetry group to be $\mbox{SO}(\N)$ instead of $\mbox{O}(\N)$ then the followings are also   superconformal  invariants in the case of even $\N$, according to Weyl~\cite{weyl}
\begin{equation}
\begin{array}{cc}
\e^{a_{1}}{}_{b_{1}}\cdots{}^{a_{\N/2}}{}_{b_{\N/2}}\displaystyle{\prod^{\frac{\N}{2}}_{j=1}}
{\cal T}_{ja_{j}}{}^{b_{j}}\,,~~~&~~~{\cal T}_{ja_{j}}{}^{b_{j}}=\hat{\bar{\Theta}}_{1(j_{1})a_{j}}\hat{\Theta}_{1(j_{2})}^{b_{j}}~~~\mbox{or}~~~
\hat{\bar{\Theta}}_{1(j_{1})a_{j}}\hat{X}_{1(j_{2})}{\cdot\gamma}\hat{\Theta}_{1(j_{3})}^{b_{j}}\,,
\end{array}
\end{equation}
which we may call pseudo-invariants.

\section{Superconformally Covariant Operators\label{operator}}
In general acting on a quasi-primary superfield, $\Psi_{\rho}{}^{r}(z)$, 
with the spinor derivative, $D_{a\alpha}$, does not lead to a
quasi-primary field\footnote{For conformally covariant differential 
operators in  non-supersymmetric theories, see e.g. \cite{9704108,9708040}.}.  
For a superfield, $\Psi_{\rho}{}^{r}$, from
eqs.(\ref{comDL},\,\ref{Ddelta}\,,\ref{deltaPsi})  we have
\begin{equation}
\begin{array}{ll}
D_{a\alpha}\delta\Psi_{\rho}{}^{r}=&
-({\cal L}+(\eta+\textstyle{\frac{1}{2}})\hat{\lambda})
D_{a\alpha}\Psi_{\rho}{}^{r}\\
{}&{}\\
{}&{}-D_{a\beta}\Psi_{\rho}{}^{r}\hat{w}^{\beta}{}_{\alpha}
-D_{a\alpha}\Psi_{\sigma}{}^{r}(s^{\beta}{}_{\gamma}
\hat{w}^{\gamma}{}_{\beta})^{\sigma}{}_{\rho}\\
{}&{}\\
{}&{}-D_{b\alpha}\Psi_{\rho}{}^{r}\hat{t}^{b}{}_{a}
-D_{a\alpha}\Psi_{\rho}{}^{s}
\textstyle{\frac{1}{2}}(s_{bc}\hat{t}{}^{bc})_{s}{}^{r}\\
{}&{}\\
{}&{}+
2\hat{\bar{\rho}}_{b\beta}(\Psi Y^{b\beta}{}_{a\alpha})_{\rho}{}^{r}\,,
\end{array}
\label{infDPSI}
\end{equation}
where $Y^{b\beta}{}_{a\alpha}$ is given by
\begin{equation}
Y^{b\beta}{}_{a\alpha}=2s^{b}{}_{a}\delta^{\beta}{}_{\alpha}
+\delta^{b}{}_{a}s^{\beta}{}_{\alpha}
-\eta\delta^{b}{}_{a}\delta^{\beta}{}_{\alpha}\,.
\end{equation}
To ensure that $D_{a\alpha}\Psi_{\rho}{}^{r}$ is quasi-primary it is
necessary that the terms proportional to $\hat{\bar{\rho}}$ vanish and 
this can be achieved by restricting $D_{a\alpha}\Psi_{\rho}{}^{r}$ 
to an irreducible representation of $\mbox{SO}(1,2)\times\mbox{O}({\cal N})$ 
and choosing a particular value of $\eta$ so that $\Psi Y=0$. 
The change of the scale dimension, 
$\eta\rightarrow\eta+\frac{1}{2}$,  in eq.(\ref{infDPSI}) is   also
apparent  from eq.(\ref{ReD}) 
\begin{equation}
D_{a\alpha}=\Omega(z;g)^{\frac{1}{2}}L^{-1\beta}{}_{\alpha}(z;g)U_{a}{}^{b}(z;g)D^{\prime}_{b\beta}\,.
\end{equation}
\newline
\indent  As an illustration we consider tensorial   
fields, $\Psi_{a_{1}\cdots a_{m}\alpha_{1}\cdots\alpha_{n}}$, 
which transform as
\begin{equation}
\begin{array}{ll}
\delta\Psi_{a_{1}\cdots a_{m}\alpha_{1}\cdots\alpha_{n}}
=&\displaystyle{
-({\cal L}+\eta\hat{\lambda})
\Psi_{a_{1}\cdots a_{m}\alpha_{1}\cdots\alpha_{n}}}\\
{}&{}\\
{}&-\displaystyle{\sum^{m}_{p=1}}
\Psi_{a_{1}\cdots b\cdots a_{m}\alpha_{1}\cdots\alpha_{n}}
\,\hat{t}{}^{b}{}_{a_{p}}
-\displaystyle{\sum^{n}_{q=1}}
\Psi_{a_{1}\cdots a_{m}\alpha_{1}\cdots \beta\cdots\alpha_{n}}
\,\hat{w}^{\beta}{}_{\alpha_{q}}\,.
\end{array}
\end{equation}
Note that spinorial indices and gauge indices, $\alpha,\,a$ may be   raised  
or lowered by  \\$\epsilon^{-1\alpha\beta},\,\e_{\alpha\beta},\,
\delta^{ab},\,\delta_{ab}$. \newline

\indent For $\Psi_{a_{1}\cdots a_{m}\alpha_{1}\cdots\alpha_{n}}$  we have
\begin{equation}
\begin{array}{ll}
(\Psi Y_{b}{}^{\beta}{}_{a\alpha})_{a_{1}\cdots a_{m}\alpha_{1}
\cdots\alpha_{n}}=&
-(\eta+\textstyle{\frac{1}{2}}n)\delta_{ba}\delta^{\beta}{}_{\alpha}
\Psi_{a_{1}\cdots a_{m}\alpha_{1}\cdots\alpha_{n}}\\
{}&{}\\
{}&\,+\delta^{\beta}{}_{\alpha}\,
\displaystyle{\sum_{p=1}^{m}}\,(\delta_{aa_{p}}
\Psi_{a_{1}\cdots b\cdots a_{m}\alpha_{1}\cdots\alpha_{n}}
-\delta_{ba_{p}}
\Psi_{a_{1}\cdots a\cdots a_{m}\alpha_{1}\cdots\alpha_{n}})\\
{}&{}\\
{}&\,+\delta_{ba}\,\displaystyle{\sum_{q=1}^{n}}\,\delta^{\beta}{}_{\alpha_{q}}
\Psi_{a_{1}\cdots\cdots a_{m}\alpha_{1}\cdots \alpha\cdots\alpha_{n}}\,.
\end{array}
\label{rhoY}
\end{equation}
In particular,  eq.(\ref{rhoY}) shows that the following are quasi-primary
\begin{subeqnarray}
\label{DPsi}
D_{[b(\beta}\Psi_{a_{1}\cdots a_{m}]\alpha_{1}\cdots\alpha_{n})}
~~~~~~~~~~~~~~~~~\mbox{if}~~~
\eta=m+\textstyle{\frac{1}{2}}n\,,\label{DPsi1}~~~~~\\
{}\nonumber\\
D_{[b|\beta|}\Psi_{a_{1}\cdots a_{m}]}{}^{\beta}
~~~~~~~~~~~~~~~~~~~~~~~\mbox{if}~~~
\eta=m-\textstyle{\frac{3}{2}}\,,\label{DPsi2}~~~~~~~
\end{subeqnarray}
where  $(\,),\,[\,]$ denote the usual
symmetrization,\,anti-symmetrization of the indices respectively and obviously 
eq.(\ref{DPsi}) is nontrivial if $1\leq  m+1\leq {\cal N}$. Note that due to the term containing $\delta_{aa_{p}}$ in  eq.(\ref{rhoY}) one should  
anti-symmetrize the gauge indices.\newline

\indent   Now we consider the case where  more than one spinor derivative,
$D_{a\alpha}$, act on a quasi-primary superfield. In this case, 
it is useful to note 
\begin{equation}
D_{[a[\alpha}D_{b]\beta]}=0\,,
\end{equation}
and
\begin{equation}
D_{a\alpha}\hat{\bar{\rho}}_{b\beta}=-i
\delta_{ab}(\e b{\cdot\gamma})_{\alpha\beta}\,.
\end{equation}
From eq.(\ref{rhoY}) one can derive 
\begin{equation}
\begin{array}{l}
D_{[b_{1}(\beta_{1}}\cdots D_{b_{l}\beta_{l}}\delta
\Psi_{a_{1}\cdots a_{m}]\alpha_{1}\cdots\alpha_{n})}\\
{}\\
=2l(-\eta+m+\textstyle{\frac{1}{2}}n+\textstyle{\frac{3}{4}}(l-1))
\hat{\bar{\rho}}_{[b_{1}(\beta_{1}}
D_{b_{2}\beta_{2}}\cdots D_{b_{l}\beta_{l}}
\Psi_{a_{1}\cdots a_{m}]\alpha_{1}\cdots\alpha_{n})}
\,+\,\mbox{homogeneous~terms}\,.
\end{array}
\end{equation}
Hence  the following is quasi-primary
\begin{equation}
\begin{array}{cc}
\label{DDPsi}
D_{[b_{1}(\beta_{1}}\cdots D_{b_{l}\beta_{l}}
\Psi_{a_{1}\cdots a_{m}]\alpha_{1}\cdots\alpha_{n})}
~~~~&~~~~\mbox{if}~~~
\eta=m+\textstyle{\frac{1}{2}}n+\textstyle{\frac{3}{4}}(l-1)\,.
\end{array}
\end{equation}
\newline
\newline

\begin{center}
\large{\textbf{Acknowledgments}}
\end{center}
I am deeply indebted to  Hugh Osborn for introducing me  the subject  
in this paper.

\newpage
\appendix
\begin{center}
\Large{\textbf{Appendix}}
\end{center}
\section{Useful Equations\label{AppendixA}}
Some useful identities relevant to the present paper  are
\begin{subeqnarray}
\label{L1}
&\textstyle{\frac{1}{2}}\mbox{tr}(\gamma^{\mu}\gamma^{\nu})
=\eta^{\mu\nu}\,,&\label{Tr2}\\
{}\nonumber\\
&\gamma^{\mu}\gamma^{\nu}\gamma^{\rho}=\eta^{\mu\nu}\gamma^{\rho}-\eta^{\mu\rho}\gamma^{\nu}+\eta^{\nu\rho}\gamma^{\mu}+i\epsilon^{\mu\nu\rho}\,,&
\label{ggg}\\
{}\nonumber\\
&-i\textstyle{\frac{1}{2}}\epsilon^{\mu\nu\rho}\gamma_{\nu}\gamma_{\rho}
=\gamma^{\mu}\,.&
\end{subeqnarray}
\begin{subeqnarray}
\label{L2}
&\e^{\mu\nu\kappa}\e_{\lambda\rho\kappa}
=\delta^{\mu}{}_{\lambda}\delta^{\nu}{}_{\rho}
-\delta^{\mu}{}_{\rho}\delta^{\nu}{}_{\lambda}\,,&\label{ee1}\\
{}\nonumber\\
&\e^{\mu\nu\kappa}\e_{\lambda\nu\kappa}=2\delta^{\mu}{}_{\lambda}\,,&
\label{ee2}\\
{}\nonumber\\
&\e^{\mu\nu\kappa}=\e_{\mu\nu\kappa}\,.&
\end{subeqnarray}
\begin{subeqnarray}
\label{L3}
&\rho\bar{\varepsilon}=-\textstyle{\frac{1}{2}}(\bar{\varepsilon}\gamma^{\mu}\rho\gamma_{\mu}+\bar{\varepsilon}\rho\,1)\,,&\label{completere}\\
{}\nonumber\\
&\e^{-1}\bar{\varepsilon}^{t}\rho^{t}\e=-\textstyle{\frac{1}{2}}
(\bar{\varepsilon}\gamma^{\mu}\rho\gamma_{\mu}-\bar{\varepsilon}\rho\,1)\,.&
\end{subeqnarray}

For  Majorana spinors
\begin{subeqnarray}
\label{L4}
&D_{a\alpha}\theta^{b\beta}=-\delta_{a}{}^{b}\delta_{\alpha}{}^{\beta}\,,~~~~~~~~~
D_{a\alpha}\bt_{b\beta}=\delta_{ab}\e_{\alpha\beta}\,,&\\
{}\nonumber\\
&\bar{D}^{a\alpha}\theta^{b\beta}=-\delta^{ab}\e^{-1\alpha\beta}\,,~~~~~~
\bar{D}^{a\alpha}\bt_{b\beta}=\delta^{a}{}_{b}\delta^{\alpha}{}_{\beta}\,,&\\
{}\nonumber\\
&D_{a\alpha}\bt_{b}\theta^{b}=2\bt_{a\alpha}\,,&\\
{}\nonumber\\
&D_{a\alpha}\x_{+}{}^{\beta}{}_{\gamma}=2i\delta_{\alpha}{}^{\beta}
\bt_{a\gamma}\,,&\\
{}\nonumber\\
&D_{a\alpha}\x_{-}{}^{\beta}{}_{\gamma}=2i(\delta_{\alpha}{}^{\beta}
\bt_{a\gamma}-\delta^{\beta}{}_{\gamma}\bt_{a\alpha})\,.&
\end{subeqnarray}
\newline
\begin{subeqnarray}
\label{L5}
&\gamma^{0}\x_{\pm}\gamma^{0}=\x_{\mp}^{\dagger}\,,&\\
{}\nonumber\\
&\e\x_{\pm}\e^{-1}=-\x_{\mp}^{t}\,,&\\
{}\nonumber\\
&\det\x_{+}=\det\x_{-}=-x^{2}-\textstyle{\frac{1}{4}}(\bt_{a}\theta^{a})^{2}\,.&
\end{subeqnarray}
\newline
\begin{equation}
\delta_{\alpha}{}^{\delta}\delta_{\beta}{}^{\gamma}-
\delta_{\alpha}{}^{\gamma}\delta_{\beta}{}^{\delta}
=\e_{\alpha\beta}\e^{-1}{}^{\gamma\delta}\,.
\label{ddcc}
\end{equation}

\begin{subeqnarray}
\label{N=1sigma2}
&\bar{\theta}\gamma^{\mu}\theta^{\prime}\bar{\psi}\gamma_{\mu}\psi^{\prime}=
-2\bt\psi^{\prime}\,\bar{\psi}\theta^{\prime}-
\bar{\theta}\theta^{\prime}\,\bar{\psi}\psi^{\prime}\,,&\\
&{}&\nonumber\\
&\epsilon_{\mu_{1}\cdots\mu_{d}}\epsilon_{\nu_{1}\cdots\nu_{d}}=\displaystyle{\sum_{p=1}^{d!}{}\mbox{sign}(p)\,
\delta_{\mu_{1}\nu_{p_{1}}}\cdots\delta_{\mu_{d}\nu_{p_{d}}}}~~~~~p:\,\mbox{permutations}\,,&\\
&{}&\nonumber\\
&\epsilon_{\mu_{1}\cdots\mu_{d}}x_{(1)}^{\mu_{1}}\cdots x_{(d)}^{\mu_{d}}=\pm\sqrt{\epsilon^{i_{1}\cdots i_{d}}x_{(1)}{\cdot x_{(i_{1})}}\cdots x_{(d)}{\cdot x_{(i_{d})}}}\,.&\label{detx}
\end{subeqnarray}

\section{Solution of Superconformal Killing Equation\label{Appendixsolving}}
From the well known  solution of the ordinary conformal Killing equation~(\ref{ordiKi})~\cite{ginsparg},  we may write the general solution of the superconformal Killing equation~(\ref{Killing1}) as 
\begin{equation}
\begin{array}{ll}
h^{\mu}(z)=&2x{\cdot b(\theta)}\,x^{\mu}-(x^{2}-
\textstyle{\frac{1}{4}}(\bt_{a}\theta^{a})^{2})b^{\mu}(\theta)+
\e^{\mu}{}_{\nu\lambda}x^{\nu}b^{\lambda}(\theta)\bt_{a}\theta^{a}\\
{}&{}\\
{}&\,+w^{\mu}{}_{\nu}(\theta)x^{\nu}+\textstyle{\frac{1}{4}}
\epsilon^{\mu}{}_{\nu\lambda}w^{\nu\lambda}(\theta)\bt_{a}\theta^{a}+\lambda(\theta) x^{\mu}+a^{\mu}(\theta)\,.
\end{array}
\label{ansatz}
\end{equation}
Substituting this expression into eq.(\ref{Killing1})  leads three independent equations corresponding to the second, first and zeroth order in $x$. Considering the quadratic terms or the coefficients of $x_{\rho}x_{\lambda}$, we get
\begin{equation}
\begin{array}{l}
\eta^{\mu\rho}D_{a\alpha}b^{\lambda}(\theta)+\eta^{\mu\lambda}D_{a\alpha}b^{\rho}(\theta)-\eta^{\rho\lambda}D_{a\alpha}b^{\mu}(\theta)\\
{}\\
=-i\textstyle{\frac{1}{2}}\e^{\mu}{}_{\nu\kappa}\gamma^{\kappa}{}^{\beta}{}_{\alpha}(
\eta^{\nu\rho}D_{a\alpha}b^{\lambda}(\theta)+\eta^{\nu\lambda}D_{a\alpha}b^{\rho}(\theta)-\eta^{\rho\lambda}D_{a\alpha}b^{\nu}(\theta))\,.
\end{array}
\label{secondx}
\end{equation}
Contracting this with $\eta_{\rho\lambda}$ gives
\begin{equation}
D_{a\alpha}b^{\mu}(\theta)=-i\textstyle{\frac{1}{2}}\e^{\mu}{}_{\nu\kappa}\gamma^{\kappa}{}^{\beta}{}_{\alpha}D_{a\beta}b^{\nu}(\theta)\,,
\end{equation}
while contraction with $\eta_{\mu\lambda}$ leads
\begin{equation}
D_{a\alpha}b^{\rho}(\theta)=
i\textstyle{\frac{1}{3}}\e^{\rho}{}_{\nu\kappa}\gamma^{\kappa}{}^{\beta}{}_{\alpha}D_{a\beta}b^{\nu}(\theta)\,.
\end{equation}
Thus
\begin{equation}
D_{a\alpha}b^{\mu}(\theta)=0\,,
\end{equation}
$b^{\mu}(\theta)$ is constant.  
Straightforward calculation shows that $2x{\cdot b}\,x^{\mu}-(x^{2}-
\textstyle{\frac{1}{4}}(\bt_{a}\theta^{a})^{2})b^{\mu}+\e^{\mu}{}_{\nu\lambda}x^{\nu}b^{\lambda}\bt_{a}\theta^{a}$ is a solution of the superconformal Killing equation~(\ref{Killing1}).\newline

\indent Now the linear in $x$ terms become
\begin{equation}
D_{a\alpha}(\e^{\mu\nu\kappa}v_{\kappa}(\theta)+\eta^{\mu\nu}\lambda(\theta))
=i\textstyle{\frac{1}{2}}D_{\a\beta}(\eta^{\mu\nu}v(\theta){\cdot\gamma}-v^{\mu}(\theta)\gamma^{\nu}-\e^{\mu\nu}{}_{\kappa}\lambda(\theta)\gamma^{\kappa})^{\beta}{}_{\alpha}\,,
\label{linearx}
\end{equation}
where $v^{\kappa}(\theta)$ is the dual form of $w^{\mu\nu}(\theta)$
\begin{equation}
\begin{array}{cc}
v^{\kappa}=\textstyle{\frac{1}{2}}\e^{\kappa}{}_{\mu\nu}w^{\mu\nu}\,,~~~~~~&~~~~~~
w^{\mu\nu}=\e^{\mu\nu\kappa}v_{\kappa}\,.
\end{array}
\end{equation}
Contracting eq.(\ref{linearx}) with $\eta_{\mu\nu}$ gives
\begin{equation}
\begin{array}{cc}
D_{a\alpha}\lambda(\theta)=i\textstyle{\frac{1}{3}}D_{a\beta}
\v^{\beta}{}_{\alpha}(\theta)\,,~~~~&~~~~\v^{\beta}{}_{\alpha}(\theta)
=v^{\mu}(\theta)\gamma_{\mu}{}^{\beta}{}_{\alpha}\,.
\end{array}
\label{lambda}
\end{equation}
Substituting this back into eq.(\ref{linearx}) leads
\begin{equation}
0=5\e^{\mu\nu\rho}D_{a\alpha}v_{\rho}(\theta)
-i\eta^{\mu\nu}D_{a\beta}\v^{\beta}{}_{\alpha}(\theta)
+4iD_{a\beta}v^{\mu}(\theta)\gamma^{\nu}{}^{\beta}{}_{\alpha}
-iD_{a\beta}v^{\nu}(\theta)\gamma^{\mu}{}^{\beta}{}_{\alpha}\,.
\label{linearx2}
\end{equation}
Contraction with $\e^{\kappa}{}_{\mu\nu}$ shows that $v^{\mu}(\theta)$ satisfies the superconformal Killing equation~(\ref{Killing1})
\begin{equation}
D_{a\alpha}v^{\kappa}(\theta)=-i\textstyle{\frac{1}{2}}\e^{\kappa}{}_{\mu\nu}
D_{a\beta}v^{\mu}(\theta)\gamma^{\nu}{}^{\beta}{}_{\alpha}\,.
\label{Killingv}
\end{equation}
Eqs.(\ref{lambda},\ref{Killingv}) are actually equivalent to eq.(\ref{linearx}), since from eq.(\ref{Killingv}) successively
\begin{equation}
\begin{array}{c}
D_{a\alpha}v^{\kappa}(\theta)\gamma^{\rho}{}^{\alpha}{}_{\beta}=
i\textstyle{\frac{1}{2}}\e^{\kappa\rho\mu}D_{a\beta}v_{\mu}(\theta)+
\textstyle{\frac{1}{2}}\eta^{\kappa\rho}D_{a\alpha}\v^{\alpha}{}_{\beta}(\theta)
-\textstyle{\frac{1}{2}}D_{a\alpha}v^{\rho}(\theta)\gamma^{\kappa}{}^{\alpha}{}_{\beta}\,,\\
{}\\
D_{a\alpha}v^{(\kappa}(\theta)\gamma^{\rho)}{}^{\alpha}{}_{\beta}=
\textstyle{\frac{1}{3}}\eta^{\kappa\rho}D_{a\alpha}\v^{\alpha}{}_{\beta}(\theta)\,,\\
{}\\
D_{a\alpha}v^{[\kappa}(\theta)\gamma^{\rho]}{}^{\alpha}{}_{\beta}=
i\e^{\kappa\rho\mu}D_{a\beta}v_{\mu}(\theta)\,,~~~~~~~~\\
{}\\
D_{a\alpha}v^{\kappa}(\theta)\gamma^{\rho}{}^{\alpha}{}_{\beta}=
i\e^{\kappa\rho\mu}D_{a\beta}v_{\mu}(\theta)+
\textstyle{\frac{1}{3}}\eta^{\kappa\rho}D_{a\alpha}\v^{\alpha}{}_{\beta}(\theta)\,,
\end{array}
\end{equation}
and the last expression makes eq.(\ref{linearx2}) hold.\newline

\indent To solve eq.(\ref{Killingv}) we first note from 
\begin{equation}
D_{a\alpha}{\rm v}^{\beta}{}_{\gamma}(\theta)
=\textstyle{\frac{2}{3}}\delta_{\alpha}{}^{\beta}D_{a\delta}
{\rm v}^{\delta}{}_{\gamma}(\theta)-\textstyle{\frac{1}{3}}
\delta^{\beta}{}_{\gamma}D_{a\delta}{\rm v}^{\delta}{}_{\alpha}(\theta)\,,
\label{Dv,v}
\end{equation}
that 
\begin{equation}
\begin{array}{ll}
D_{b\beta}D_{a\alpha}\v^{\gamma}{}_{\delta}(\theta)&=
-\textstyle{\frac{2}{3}}\delta_{\alpha}{}^{\gamma}D_{a\omega}D_{b\beta}\v^{\omega}{}_{\delta}(\theta)+\textstyle{\frac{1}{3}}
\delta^{\gamma}{}_{\delta}D_{a\omega}D_{b\beta}
{\rm v}^{\omega}{}_{\alpha}(\theta)\\
{}&{}\\
{}&=\textstyle{\frac{4}{9}}\delta_{\alpha}{}^{\gamma}D_{b\omega}D_{a\beta}\v^{\omega}{}_{\delta}(\theta)-\textstyle{\frac{2}{9}}\delta_{\alpha}{}^{\gamma}D_{b\omega}D_{a\delta}\v^{\omega}{}_{\beta}(\theta)-\textstyle{\frac{2}{9}}\delta^{\gamma}{}_{\delta}D_{b\omega}D_{a\beta}\v^{\omega}{}_{\alpha}(\theta)\\
{}&{}\\
{}&\,\,~~+\textstyle{\frac{1}{9}}\delta^{\gamma}{}_{\delta}D_{b\omega}D_{a\alpha}
\v^{\omega}{}_{\beta}(\theta)\,.
\end{array}
\label{DDv}
\end{equation}
Contraction with $\delta_{\gamma}{}^{\beta}$ gives
\begin{equation}
D_{b\gamma}D_{a\alpha}\v^{\beta}{}_{\delta}(\theta)=-
D_{b\gamma}D_{a\delta}\v^{\beta}{}_{\alpha}(\theta)\,,
\label{DDvanti}
\end{equation}
so that eq.(\ref{DDv}) becomes
\begin{equation}
D_{b\beta}D_{a\alpha}\v^{\gamma}{}_{\delta}(\theta)=\textstyle{\frac{2}{3}}
\delta_{\alpha}{}^{\gamma}D_{b\omega}D_{a\beta}\v^{\omega}{}_{\delta}(\theta)-\textstyle{\frac{1}{3}}\delta^{\gamma}{}_{\delta}D_{b\omega}D_{a\beta}
\v^{\omega}{}_{\alpha}(\theta)\,,
\label{DDv2}
\end{equation}
which is in fact equivalent to eq.(\ref{DDv}).\newline

From eq.(\ref{DDv2}) and $D_{b\beta}D_{a\alpha}\v^{\gamma}{}_{\delta}(\theta)=-D_{a\alpha}
D_{b\beta}\v^{\gamma}{}_{\delta}(\theta)$ we get
\begin{equation}
2\delta_{\alpha}{}^{\gamma}D_{b\omega}D_{a\beta}\v^{\omega}{}_{\delta}(\theta)+
2\delta_{\beta}{}^{\gamma}D_{a\omega}D_{b\alpha}\v^{\omega}{}_{\delta}(\theta)
=\delta^{\gamma}{}_{\delta}(D_{a\omega}D_{b\alpha}\v^{\omega}{}_{\beta}(\theta)+D_{b\omega}D_{a\beta}\v^{\omega}{}_{\alpha}(\theta))\,.
\end{equation}
Contracting with $\delta_{\gamma}{}^{\alpha}$ gives
\begin{equation}
3D_{b\omega}D_{a\beta}\v^{\omega}{}_{\delta}(\theta)=
-2D_{a\omega}D_{b\beta}\v^{\omega}{}_{\delta}(\theta)
+D_{a\omega}D_{b\delta}\v^{\omega}{}_{\beta}(\theta)\,.
\end{equation}
Hence from eq.(\ref{DDvanti}) we can put
\begin{equation}
\begin{array}{l}
D_{b\omega}D_{a\alpha}\v^{\omega}{}_{\beta}=\Gamma_{ab}(\theta)
\e_{\alpha\beta}\,,\\
{}\\
\Gamma_{ab}(\theta)=\textstyle{\frac{1}{2}}D_{b\beta}D_{a\alpha}
(v(\theta)\e^{-1})^{\beta\alpha}=-\Gamma_{ba}(\theta)\,,
\end{array}
\end{equation}
so that eq.(\ref{DDv2}) becomes with eq.(\ref{ddcc})
\begin{equation}
D_{b\beta}D_{a\alpha}\v^{\gamma}{}_{\delta}(\theta)=
\textstyle{\frac{1}{3}}(2\delta_{\alpha}{}^{\gamma}\e_{\beta\delta}
-\e_{\beta\alpha}\delta^{\gamma}{}_{\delta})\Gamma_{ab}(\theta)\,.
\label{DDvG}
\end{equation}
Thus
\begin{equation}
\begin{array}{ll}
D_{c\gamma}\Gamma_{ab}(\theta)&=\textstyle{\frac{1}{2}}D_{b\beta}D_{a\alpha}D_{c\gamma}(\v(\theta)\e^{-1})^{\beta\alpha}\\
{}&{}\\
{}&=\textstyle{\frac{1}{2}}D_{b\gamma}\Gamma_{ca}(\theta)\\
{}&{}\\
{}&=0\,.
\end{array}
\end{equation}
Therefore $\Gamma_{ab}(\theta)$ is independent of $\theta$ and $\v(\theta)$ is at most quadratic in $\theta$. \newline
From eq.(\ref{DDvG}) we get
\begin{equation}
D_{b\beta}D_{a\gamma}\v^{\gamma}{}_{\delta}(\theta)=\e_{\beta\delta}\Gamma_{ab}\,.
\end{equation}
Integrating this gives
\begin{equation}
D_{a\gamma}\v^{\gamma}{}_{\alpha}(\theta)=6i(t_{a}{}^{b}\bt_{b\alpha}
+\bar{\rho}_{a\alpha})\,,
\end{equation}
where $6it_{a}{}^{b}=\Gamma_{ab}$ so that 
\begin{equation}
t^{t}=-t\,,
\label{KtK2}
\end{equation}
and the spinor, $\bar{\rho}_{a\alpha}$, appears as  a constant of integration.\\
Now eq.(\ref{Dv,v}) becomes with eq.(\ref{complete})
\begin{equation}
D_{a\alpha}\v^{\beta}{}_{\gamma}(\theta)=
2i(t_{a}{}^{b}\bt_{b}\gamma^{\mu}+\bar{\rho}_{a}\gamma^{\mu})_{\alpha}\gamma_{\mu}{}^{\beta}{}_{\gamma}\,.
\end{equation}
Integrating this  gives
\begin{equation}
v^{\mu}(\theta)=it_{a}{}^{b}\bt_{b}\gamma^{\mu}\theta^{a}
+2i\bar{\rho}_{a}\gamma^{\mu}\theta^{a}+v^{\mu}\,.
\label{solutionforv}
\end{equation}
Eq.(\ref{lambda}) becomes
\begin{equation}
D_{a\alpha}\lambda(\theta)=-2t_{a}{}^{b}\bt_{b\alpha}-2\bar{\rho}_{a\alpha}\,,
\end{equation}
so that
\begin{equation}
D_{b\beta}D_{a\alpha}\lambda(\theta)
=-i\textstyle{\frac{1}{3}}\Gamma_{ab}\e_{\alpha\beta}\,.
\end{equation}
However from $D_{b\beta}D_{a\alpha}\lambda(\theta)+D_{a\alpha}D_{b\beta}
\lambda(\theta)=0$ we note $\Gamma_{ab}=0$. Hence
\begin{subeqnarray}
&w^{\mu\nu}(\theta)=\bar{\rho}_{a}
(\gamma^{\mu}\gamma^{\nu}-\gamma^{\nu}\gamma^{\mu})\theta^{a}+w^{\mu\nu}\,,&\\
{}\nonumber\\
&\lambda(\theta)=-2\bar{\rho}_{a}\theta^{a}+\lambda\,.&
\end{subeqnarray}
With these expressions straightforward calculation shows that 
$w^{\mu}{}_{\nu}(\theta)x^{\nu}+\textstyle{\frac{1}{4}}
\epsilon^{\mu}{}_{\nu\lambda}w^{\nu\lambda}(\theta)\bt_{a}\theta^{a}+\lambda(\theta) x^{\mu}$ is a solution of the superconformal Killing equation~(\ref{Killing1}).\newline

\indent The remaining terms are
\begin{equation}
D_{a\alpha}a^{\mu}(\theta)=-i\textstyle{\frac{1}{2}}\e^{\mu}{}_{\lambda\rho}
D_{a\beta}a^{\lambda}(\theta)\gamma^{\rho}{}^{\beta}{}_{\alpha}\,,
\label{Killinga}
\end{equation}
the general solution of which we already obtained. From eq.(\ref{solutionforv})
\begin{equation}
a^{\mu}(\theta)=it_{a}{}^{b}\bt_{b}\gamma^{\mu}\theta^{a}
+2i\bar{\varepsilon}_{a}\gamma^{\mu}\theta^{a}+a^{\mu}\,.
\label{solutionfora}
\end{equation}
For $a^{\mu}(\theta)$ to be real $t$ must be anti-hermitian and hence with 
eq.(\ref{KtK2}) $t\in\mbox{o}(\N)$.\newline

\indent All together, we obtain  the general solution of 
the superconformal Killing equation~(\ref{solutionforh}).

\section{Basis for  Superconformal Algebra\label{AppendixC}}
We write the superconformal generators in general as 
\begin{equation}
{\cal K}{\cdot{\cal P}}=
a^{\mu}P_{\mu}+\bar{\varepsilon}_{a}Q^{a}+\lambda D+
\textstyle{\frac{1}{2}}w^{\mu\nu}M_{\mu\nu}+b^{\mu}K_{\mu}+
\bar{\rho}_{a}S^{a}+\textstyle{\frac{1}{2}}t^{ab}A_{ab}\,,
\end{equation}
for
\begin{subeqnarray}
&{\cal K}=(a^{\mu},b^{\mu},\varepsilon^{a},\rho^{a},
\lambda,w^{\mu\nu},t_{a}{}^{b})\,,&\\
{}\nonumber\\
&{\cal P}=(P_{\mu},K_{\mu},Q^{a},S^{a},D,M_{\mu\nu},
A_{a}{}^{b})\,,&
\end{subeqnarray}
where we put $t^{ab}=\delta^{ac}t_{c}{}^{b}$ and 
the $R$-symmetry generators, $A_{ab}=A_{a}{}^{c}\delta_{cb}$,  
satisfy the $\mbox{o}(\N)$ 
condition, $A^{\dagger}=A^{t}=-A$.\newline

\indent The  superconformal algebra can now be obtained by imposing
\begin{equation}
[{\cal K}_{1}{\cdot{\cal P}},{\cal K}_{2}{\cdot{\cal P}}]
=-i{\cal K}_{3}{\cdot{\cal P}}\,,
\end{equation}
where ${\cal K}_{3}$ is given by  eq.(\ref{MMcom}).  
From this expression, we can read off 
the following  superconformal algebra. 
\begin{itemize}
\item Poincar\'{e} algebra
\begin{equation}
\begin{array}{cc}
[P_{\mu},P_{\nu}]=0\,, &
[M_{\mu\nu},P_{\lambda}]=i(\eta_{\mu\lambda}P_{\nu}
-\eta_{\nu\lambda}P_{\mu})\,,\\
{}&{}\\
\multicolumn{2}{l}{[M_{\mu\nu},M_{\lambda\rho}]=
i(\eta_{\mu\lambda}M_{\nu\rho}-\eta_{\mu\rho}M_{\nu\lambda}
-\eta_{\nu\lambda}M_{\mu\rho}+\eta_{\nu\rho}M_{\mu\lambda})\,.~~~~~~~}
\end{array}
\label{poincare}
\end{equation}
\item Supersymmetry algebra
\begin{equation}
\begin{array}{c}
\{Q^{a\alpha}, \bar{Q}_{b\beta}\}
=2\delta^{a}{}_{b}\gamma^{\mu}{}^{\alpha}{}_{\beta}P_{\mu}\,,\\
{}\\
{}[M_{\mu\nu},Q^{a}]=i\textstyle{\frac{1}{2}}
\gamma_{[\mu}\gamma_{\nu]}Q^{a}\,,\\
{}\\
{}[{P}_{\mu},Q^{a\alpha}]=0\,.
\end{array}
\end{equation}
\item Special superconformal algebra
\begin{equation}
\begin{array}{c}
[K_{\mu},K_{\nu}]=0\,,~~~~~~~~~
[M_{\mu\nu},K_{\lambda}]=i(\eta_{\mu\lambda}K_{\nu}
-\eta_{\nu\lambda}K_{\mu})\,,\\
{}\\
\{S^{a\alpha}, \bar{S}_{b\beta}\}
=2\delta^{a}{}_{b}\gamma^{\mu}{}^{\alpha}{}_{\beta}K_{\mu}\,,\\
{}\\
{}[M_{\mu\nu},S^{a}]=i\textstyle{\frac{1}{2}}
\gamma_{[\mu}\gamma_{\nu]}S^{a}\,,\\
{}\\
{}[{K}_{\mu},S^{a\alpha}]=0\,.
\end{array}
\end{equation}
\item Cross terms between $(P,Q)$ and $(K,S)$
\begin{equation}
\begin{array}{c}
{}[P_{\mu},K_{\nu}]=2i(M_{\mu\nu}+\eta_{\mu\nu}D)\,,\\
{}\\
{}[P_{\mu},S^{a}]=-\gamma_{\mu}Q^{a}\,,\\
{}\\
{}[K_{\mu},Q^{a}]=-\gamma_{\mu}S^{a}\,,\\
{}\\
{}\{Q^{a\alpha},\bar{S}_{b\beta}\}=-i\delta^{a}{}_{b}
(2\delta^{\alpha}{}_{\beta}D+
(\gamma^{[\mu}\gamma^{\nu]})^{\alpha}{}_{\beta}M_{\mu\nu})
+2i\delta^{\alpha}{}_{\beta}A^{a}{}_{b}\,.
\end{array}
\end{equation}
\item Dilations
\begin{equation}
\begin{array}{ll}
{}~~~[D,P_{\mu}]=-iP_{\mu}\,,~~~~&~~~~[D,K_{\mu}]=iK_{\mu}\,,\\
{}&{}\\
{}~~~[D,Q^{a}]=-i\textstyle{\frac{1}{2}}Q^{a}\,,~~~~&~~~~
{[D,S^{a}]=i\textstyle{\frac{1}{2}}S^{a}}\,,\\
{}&{}\\
\multicolumn{2}{c}{[D,D]=[D,M_{\mu\nu}]=[D,A_{a}{}^{b}]=0\,.}
\end{array}
\end{equation}
\item R-symmetry, $\mbox{o}(\N)$
\begin{equation}
\begin{array}{c}
[A_{ab},A_{cd}]
=i(\delta_{ac}A_{bd}-\delta_{ad}A_{bc}-\delta_{bc}A_{ad}+\delta_{bd}A_{ac})
\,,\\
{}\\
{}
[A_{ab},Q^{c}]=i(\delta_{a}{}^{c}\delta_{bd}-\delta_{b}{}^{c}\delta_{ad})Q^{d}\,,\\
{}\\
{}[A_{ab},S^{c}]=i(\delta_{a}{}^{c}\delta_{bd}-\delta_{b}{}^{c}\delta_{ad})
S^{d}\,,\\
{}\\
{}[A_{a}{}^{b},P_{\mu}]=[A_{a}{}^{b},K_{\mu}]=[A_{a}{}^{b},M_{\mu\nu}]=0\,.
\end{array}
\label{USU}
\end{equation}
\end{itemize}

\section{Realization of 
$\mbox{SO}(2,3)\stackrel{{}_{{}_{\displaystyle{\sim}}}}{\displaystyle{=}}
\mbox{Sp}(2,\mbox{\textsf{R}})$  structure in $M$\label{AppendixD}}
We exhibit explicitly the relation of the three-dimensional conformal 
group to \\
$\mbox{SO}(2,3)\cong\mbox{Sp}(2,\Real)$ by introducing 
five-dimensional gamma  matrices, $\Gamma^{A}$,  $A=0,1,\cdots,4$ 
\begin{equation}
\begin{array}{lll}
\Gamma^{\mu}=\left(\begin{array}{cc}
                 \gamma^{\mu} &0\\
                     0&-\gamma^{\mu}
                  \end{array}\right)\,,~~~~&~~~
\Gamma^{3}=\left(\begin{array}{cc}
                     0&i\\
                     i&0
                  \end{array}\right)\,,~~~~&~~~
\Sigma^{4}=\left(\begin{array}{cc}
                     0&i\\
                    -i&0
                  \end{array}\right)\,.
\end{array}
\end{equation}
They satisfy with $G^{AB}
=\mbox{diag}(+1,-1,-1,-1,+1)$
\begin{equation}
\Gamma^{A}\Gamma^{B}+\Gamma^{B}\Gamma^{A}=2G^{AB}\,,
\end{equation}
and
\begin{equation}
\begin{array}{cc}
\left(\begin{array}{cc}
        0&\gamma^{0}\\
         \gamma^{0}&0
        \end{array}\right)\Gamma^{A}\left(\begin{array}{cc}
        0&\gamma^{0}\\
         \gamma^{0}&0
        \end{array}\right)=-\Gamma^{A}{}^{\dagger}\,,~~~~&~~~~
\left(\begin{array}{cc}
        0&\e\\
         \e&0
        \end{array}\right)\Gamma^{A}\left(\begin{array}{cc}
        0&\e^{-1}\\
         \e^{-1}&0
        \end{array}\right)=\Gamma^{A}{}^{t}\,.
\end{array}
\label{bc}
\end{equation}
For the supermatrix,~$M$, given in eq.(\ref{Mform}), we may now express 
the $4\times 4$ part in terms of 
$\Gamma^{AB}\equiv\frac{1}{4}[\Gamma^{A},\Gamma^{B}]$ as
\begin{equation}
m\equiv\left(\begin{array}{cc}
w+\textstyle{\frac{1}{2}}\lambda &ia{\cdot\gamma}\\
ib{\cdot\gamma}&w-\textstyle{\frac{1}{2}}\lambda
\end{array}\right)
=\textstyle{\frac{1}{2}}w_{AB}\Gamma^{AB}\,,
\end{equation}
where $w_{34},\,w_{\mu 3},\,w_{\mu 4}$ are given by
\begin{equation}
\begin{array}{ccc}
w_{34}=\lambda\,,~~~~&~~~~w_{\mu 3}=a_{\mu}-b_{\mu}\,,~~~~&~~~~
w_{\mu 4}=a_{\mu}+b_{\mu}\,.
\end{array}
\end{equation}
$\Gamma^{AB}$ 
generates the Lie algebra of $\mbox{SO}(2,3)$
\begin{equation}
[\Gamma^{AB},\Gamma^{CD}]=
-G^{AC}\Gamma^{BD}+G^{AD}\Gamma^{BC}
+G^{BC}\Gamma^{AD}-G^{BD}\Gamma^{AC}\,.
\end{equation}
In general, $m$ can be defined as a  
$4\times 4$ matrix subject to two conditions
\begin{subeqnarray}
\label{mcon}
&bm+m^{\dagger}b=0\,,~~~~~~~~~
b=\left({\begin{array}{cc}
        0&\gamma^{0}\\
         \gamma^{0}&0
        \end{array}}\right)\,,&\\
{}\nonumber\\
&cm+m^{t}c=0\,,~~~~~~~~~
c=\left({\begin{array}{cc}
        0&\e\\
         \e&0
        \end{array}}\right)\,,~~~&
\end{subeqnarray}

\indent To show $\mbox{SO}(2,3)\cong \mbox{Sp}(2,\Real)$ we 
take, without loss of generality, $\gamma^{0}=i\e$ and $\e$ to be real. 
Now if we define
\begin{equation}
\begin{array}{cc}
\tilde{m}=pmp^{-1}\,,~~~~&~~~~
p=\left(\begin{array}{cc}
1&0\\
0&\e
\end{array}\right)\,,
\end{array}
\end{equation}
then from
\begin{equation}
\begin{array}{cc}
p^{-1}=p^{\dagger}=p^{t}\,,~~~~&~~~~
pcp^{-1}=\left(\begin{array}{cc}
0&1\\
-1&0\end{array}\right)=j\,,
\end{array}
\end{equation}
we note that eq.(\ref{mcon}) is equivalent to the $\mbox{sp}(2,\Real)$ 
condition
\begin{equation}
\begin{array}{cc}
\tilde{m}^{\ast}=\tilde{m}\,,~~~~&~~~~
j\tilde{m}+\tilde{m}^{t}j=0\,.
\end{array}
\end{equation}

\bibliographystyle{unsrt}
\bibliography{reference}

\end{document}